\documentclass[sigplan,10pt]{acmart}
\acmConference[PPoPP'26]{ACM SIGPLAN Annual Symposium on Principles and Practice of Parallel Programming}{January 31--February 04, 2026}{Sydney, Australia}
\acmYear{2026}
\acmISBN{} 
\acmDOI{} 
\startPage{1}
\setcopyright{none}
\usepackage[utf8]{inputenc}
\usepackage{hyperref}
\usepackage{algorithmic}
\usepackage{graphicx}
\usepackage{textcomp}
\usepackage{xcolor}
\usepackage{soul}
\usepackage{algorithm}
\usepackage{fancyvrb}
\usepackage{pifont}
\usepackage{subcaption}
\VerbatimFootnotes
\usepackage{booktabs}
\usepackage{url}
\newcommand{\RR}{\mathbb{R}}
\usepackage{microtype}
\usepackage{todonotes}
\usepackage{multirow}

\renewcommand\footnotetextcopyrightpermission[1]{} 

\usepackage{caption}
\usepackage{enumitem}
\setlist{nosep}

\usepackage{amsmath}

\usepackage{diagbox}
\usepackage{amsfonts}

\begin{abstract}
Large symmetric eigenvalue problems are commonly observed in many disciplines such as Chemistry and Physics, and several libraries including cuSOLVERMp, MAGMA and ELPA support computing large eigenvalue decomposition on multi-GPU or multi-CPU-GPU hybrid architectures. However, these libraries do not provide satisfied performance that all of the libraries only utilize around 1.5\% of the peak multi-GPU performance. In this paper, we propose a pipelined two-stage eigenvalue decomposition algorithm instead of conventional subsequent algorithm with substantial optimizations. 
On an 8$\times$A100 platform, our implementation surpasses state-of-the-art cuSOLVERMp and MAGMA baselines, delivering mean speedups of 5.74$\times$ and 6.59$\times$, with better strong and weak scalability.

\end{abstract}

\begin{document}

\title{Pipelined Dense Symmetric Eigenvalue Decomposition on Multi-GPU Architectures}

\author{Hansheng Wang}
\affiliation{University of Electronic Science and Technology of China}
\email{wanghansheng@std.uestc.edu.cn}

\author{Ruiyi Zhan}
\affiliation{University of Electronic Science and Technology of China}
\email{rui1@std.uestc.edu.cn}

\author{Dajun Huang}
\affiliation{University of Electronic Science and Technology of China}
\email{djhuang@std.uestc.edu.cn}

\author{Xingchen Liu}
\affiliation{University of Chinese Academy of Sciences}
\email{liuxingchen232@mails.ucas.ac.cn}

\author{Qiao Li}
\affiliation{Xiamen University}
\email{qiaoli045@gmail.com}

\author{Hancong Duan}
\affiliation{University of Electronic Science and Technology of China}
\email{duanhancong@uestc.edu.cn}

\author{Dingwen Tao}
\affiliation{Institute of Computing Technology, Chinese Academy of Sciences}
\email{taodingwen@ict.ac.cn}

\author{Guangming Tan}
\affiliation{Institute of Computing Technology, Chinese Academy of Sciences}
\email{tgm@ict.ac.cn}

\author{Shaoshuai Zhang}
\affiliation{University of Electronic Science and Technology of China}
\email{szhang94@uestc.edu.cn}

\maketitle
\pagestyle{plain}

\section{Introduction}

Eigenvalue decomposition (EVD) aims to factorizes a matrix to a diagonal form: $
A=Q \times \Lambda \times Q^T, A\in \RR^{n\times n},
$
where $\Lambda$ is a diagonal matrix contains the eigenvalues on its diagonal, and $Q$ is an orthogonal matrix consists of eigenvectors. Given a dense symmetric matrix, the entire EVD process typically include three subsequent processes: tridiagonalization~\cite{blockReduction}, iterative solver~\cite{QRAlgorithm, DivideAndConquer} and back transformation~\cite{2stage1}. Among the three processes, tridiagonalization reduces the full matrix to a tridiagonal form; the iterative solver such as QR algorithm~\cite{QRAlgorithm} and Divide and Conquer (D\&C)~\cite{DivideAndConquer} further diagonalize the tridiagonal matrix and generates the eigenvalues. If the eigenvectors are needed, then the iterative solver also generates the eigenvectors of the tridiagonal matrix, and the back transformation process will form the final eigenvectors. 

In dense symmetric EVD, two principal algorithmic families are widely used: one-stage EVD~\cite{blockReduction} and two-stage EVD~\cite{2stage1,2stage2,2stage3}. The one-stage approach directly applies Householder reflectors to reduce the original matrix to tridiagonal form. 
However, this approach relies heavily on BLAS2 operations, which exhibit low arithmetic intensity and suboptimal data reuse. Consequently, it fails to exploit modern GPUs' computing capacity.

To mitigate these inefficiencies, the two-stage approach decomposes tridiagonalization into two phases: successive band reduction (SBR) first reduces the dense matrix to a banded form, and bulge chasing (BC) then converts the band matrix to tridiagonal matrix. By recasting the bulk of the computation into BLAS3 operations, two-stage tridiagonalization achieves up to 10x speedup over one-stage EVD~\cite{DBBR}. Thus, this paper focuses on optimizing the two-stage EVD.

In real-world applications, including tight-binding models in condensed matter physics~\cite{tightBinding}, quantum chemistry~\cite{quantumChemistry1}, and density functional theory problems~\cite{densityFunction}, scientists often need to solve large EVD problems that exceed the memory capacity and computational efficiency limits of a single GPU. Consequently, multi-GPU EVD solvers have been developed. Among these, cuSOLVERMp~\footnote{\url{https://docs.nvidia.com/cuda/cusolvermp/}} employs a one-stage EVD approach, while libraries such as ELPA~\cite{marek2014elpa}, MAGMA~\cite{MAGMA}, and SLATE~\cite{gates2020slate} support both one-stage and two-stage EVD methods.

However, as shown in Table~\ref{tbl:perf_evd_A100}, given a $49152\times 49152$ matrix, even the state-of-the-art (SOTA) multi-GPU EVD libraries such as MAGMA and cuSOLVERMp achieve only 2.18 TFLOPS and 2.37 TFLOPS on 8 A100 GPUs, respectively. These amounts only reach 1.3\% and 1.5\% of the GPUs' peak performance ($8\times19.5$ peek TFLOPS), highlighting a significant performance gap in current solutions.

In this paper, we suggest the performance bottlenecks of existing multi-GPU two-stage EVD implementations mainly lie in the underutilization of GPU resources, and we thereby propose a pipelined EVD algorithm. The pipeline introduces parallelism across consecutive stages, reduces synchronization overhead and improves GPU utilization. Compared to the SOTA implementations in cuSOLVERMp and MAGMA, the proposed EVD algorithm achieves significant performance improvements and better scalability, demonstrating superior efficiency in large-scale eigenvalue computations on modern multi-GPU architectures.

We consider the contributions of this paper to be:
\begin{itemize}

\item We introduce a new pipeline scheme for EVD solvers to fully utilize the GPU resources.
\item We design a blockwise data allocation strategy instead of block-cyclic strategy to satisfy the requirement of pipeline, and propose a new load balance approach to provide overall load balance in the pipeline.
\item We further optimize the tridiagonalization and back transformation process to reduce the communication volume and improve the kernel performance on GPU architectures.
\item We conduct sufficient experiments on the proposed EVD algorithm to show that it has better performance and scalability than the SOTA multi-GPU implementations.

\end{itemize}

The remainder of the paper is organized as follows: Section 2 introduces the fundamental concepts of EVD. Section 3 demonstrates the motivation, difficulties and solutions of designing pipelined EVD. Section 4 details our implementation and optimizations. Section 5 presents experimental results, while Section 6 discusses related work. Finally, Section 7 concludes the paper and outlines future work.

\begin{table}[htbp]
    \centering
    \setlength{\tabcolsep}{3.5pt} 
    \begin{tabular}{c cccc}
  \toprule
  $n$ & \multicolumn{2}{c}{Time} & \multicolumn{2}{c}{FLOPS} \\
  \cmidrule(lr){2-3} \cmidrule(lr){4-5}
      & MAGMA & cuSOLVERMp & MAGMA & cuSOLVERMp \\
  \midrule
  49152 & 217.53s & 200.02s & $2.18T$ & $2.37T$ \\
  \bottomrule
  \end{tabular}%
   \caption{Performance comparison of MAGMA and cuSOLVERMp on $8\times A100$ GPUs for matrix size $n=49152$, with TFLOPS computed as $\frac{4n^3}{\text{Time}}$.}\label{tbl:perf_evd_A100}
\end{table}

\section{Background}
\subsection{Two-Stage Eigenvalue Decomposition}

\subsubsection{Full to Diagonal Steps}

\begin{figure}
    \centering
    \includegraphics[width=0.95\columnwidth]{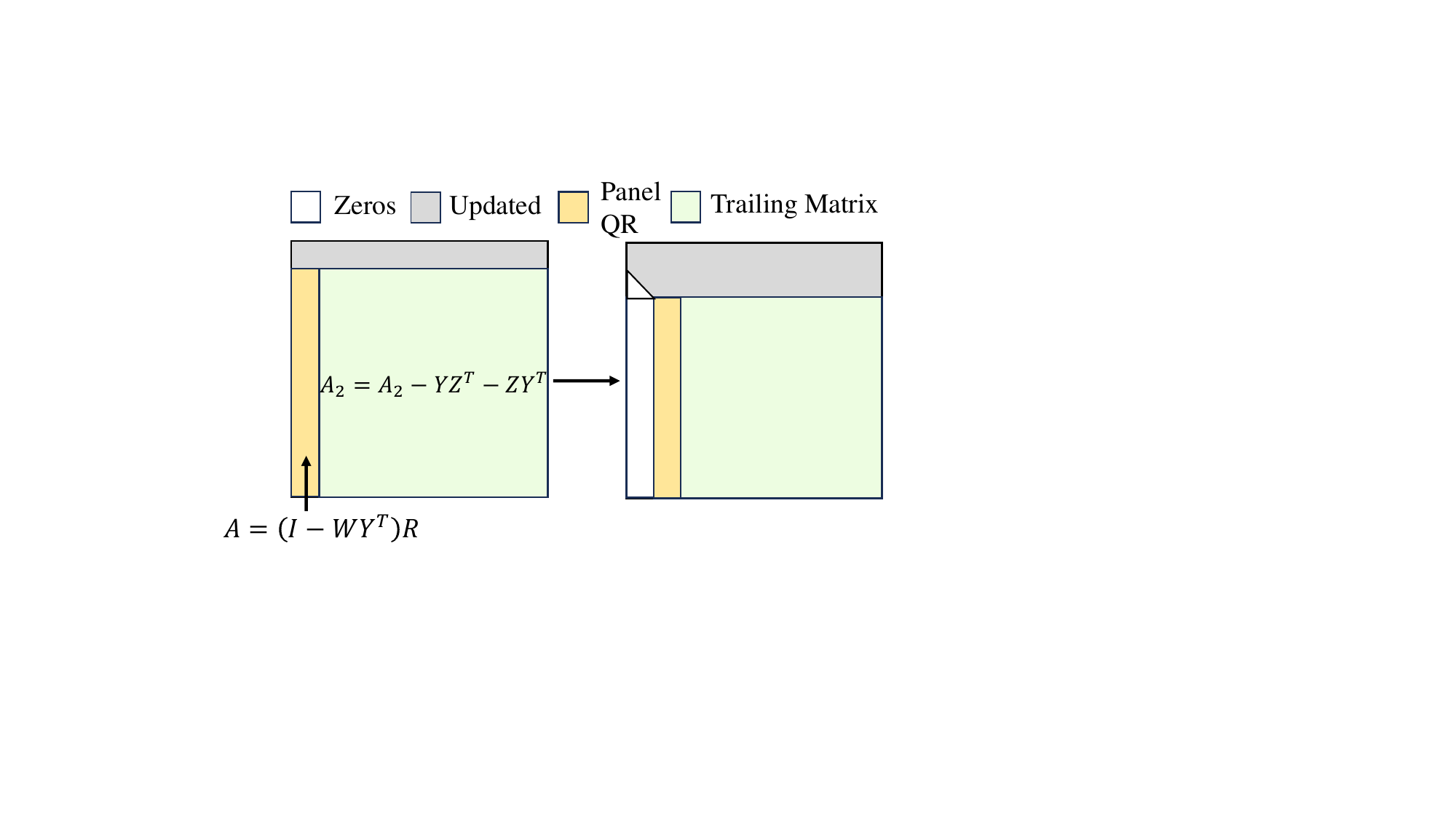}
    \caption{The first two iterations in SBR~\cite{DBBR}.\label{fig:SBR_process}}
\end{figure}

There're three consecutive steps to convert a full matirx to a diagonal matrix in two-stage EVD: successive band reduction (SBR), bulge chasing (BC) and iterative solver.

Figure~\ref{fig:SBR_process} demonstrates the initial two stages of the SBR process, which transforms the original matrix into a band matrix. The algorithm begins by selecting a panel (light yellow block) for QR factorization, resulting in the decomposition $QR(\text{Panel}) = (I - WY^T)R$. This factorization eliminates off-band elements, with the $R$ matrix replacing the upper triangular portion of the panel. The subsequent trailing matrix update employs a two-sided operation using the ZY representation, which efficiently utilizes the \verb|syr2k| routine~\cite{blockReduction} as shown in Equation~\ref{eqa: ZY1}.

\begin{equation}
\begin{aligned}
    & Z=AW-\frac{1}{2}YW^TAW\\
   & A_2=A_2-YZ^T-ZY^T
    \end{aligned}
    \label{eqa: ZY1}
\end{equation}

Following this update, the modified trailing matrix (now a new full matrix) becomes the input for the next iteration of the SBR process, enabling a recursive reduction of the original matrix to band form.

After the SBR process, BC will be executed to convert the band matrix to a triangular matrix using the similar Householder transformation steps in SBR, but its time complexity is much lower ($O(n^3)$ versus $O(n^2b)$, where $b$ is bandwidth of the band form matrix), and its rich of BLAS2 operations, more details can be found in~\cite{DBBR}.

When BC is finished, the tridiagonal matrix $T$ will be  obtained, the iterative solvers such as D\&C~\cite{DivideAndConquer} and QR algorithm~\cite{QRAlgorithm} will be performed to diagonalize the matrix $T$ to $\Lambda$ with the eigenvalues presented on its diagonal.

\subsubsection{Back Transformation}
\begin{figure}
    \centering
    \includegraphics[width=1.0\columnwidth]{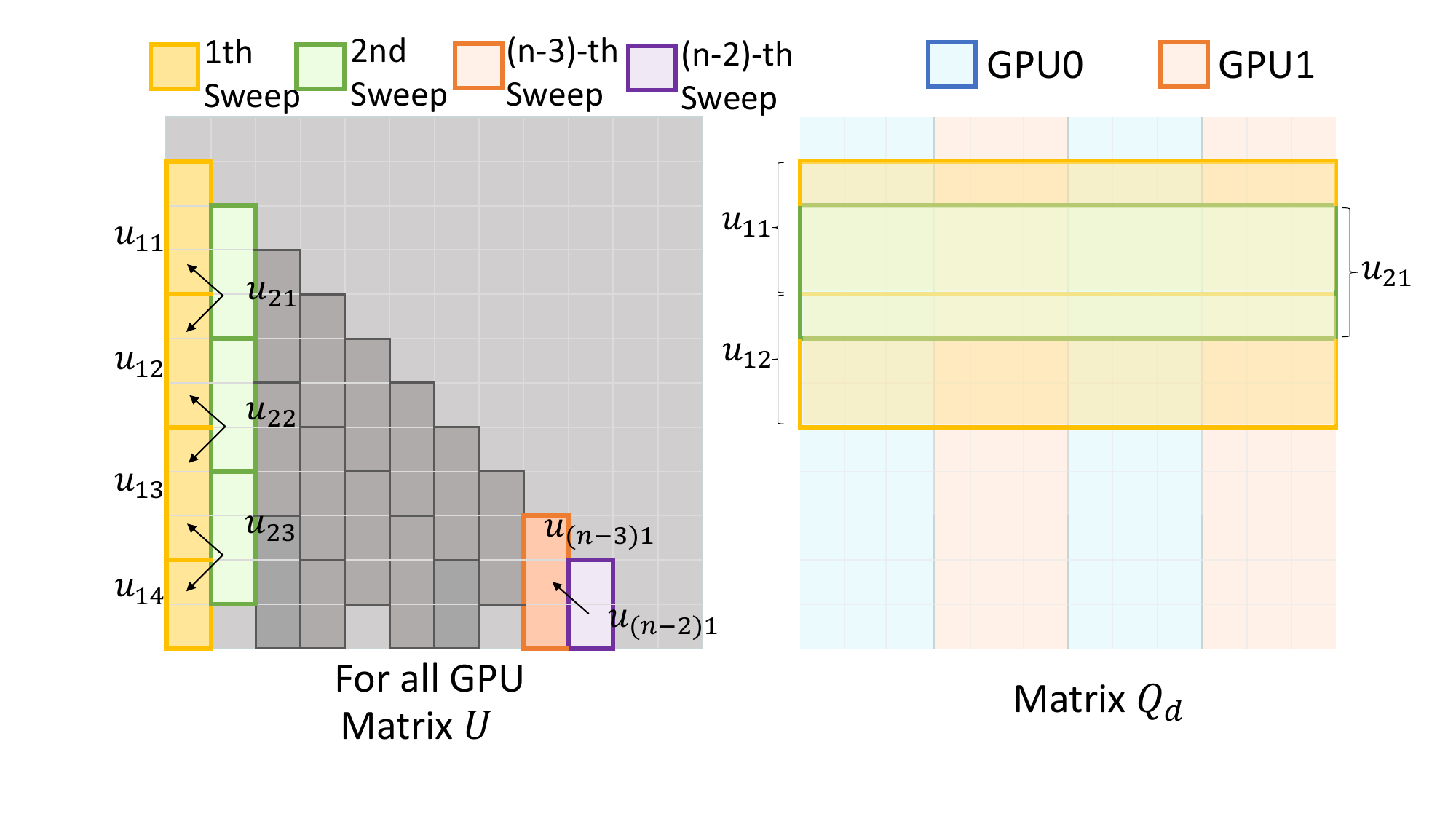}
    \caption{The illustration of BC-Back transformation multiplies $Q_d$.\label{fig:backtrans_blas2}}
\end{figure}

\begin{figure}
    \centering
    \includegraphics[width=1\linewidth]{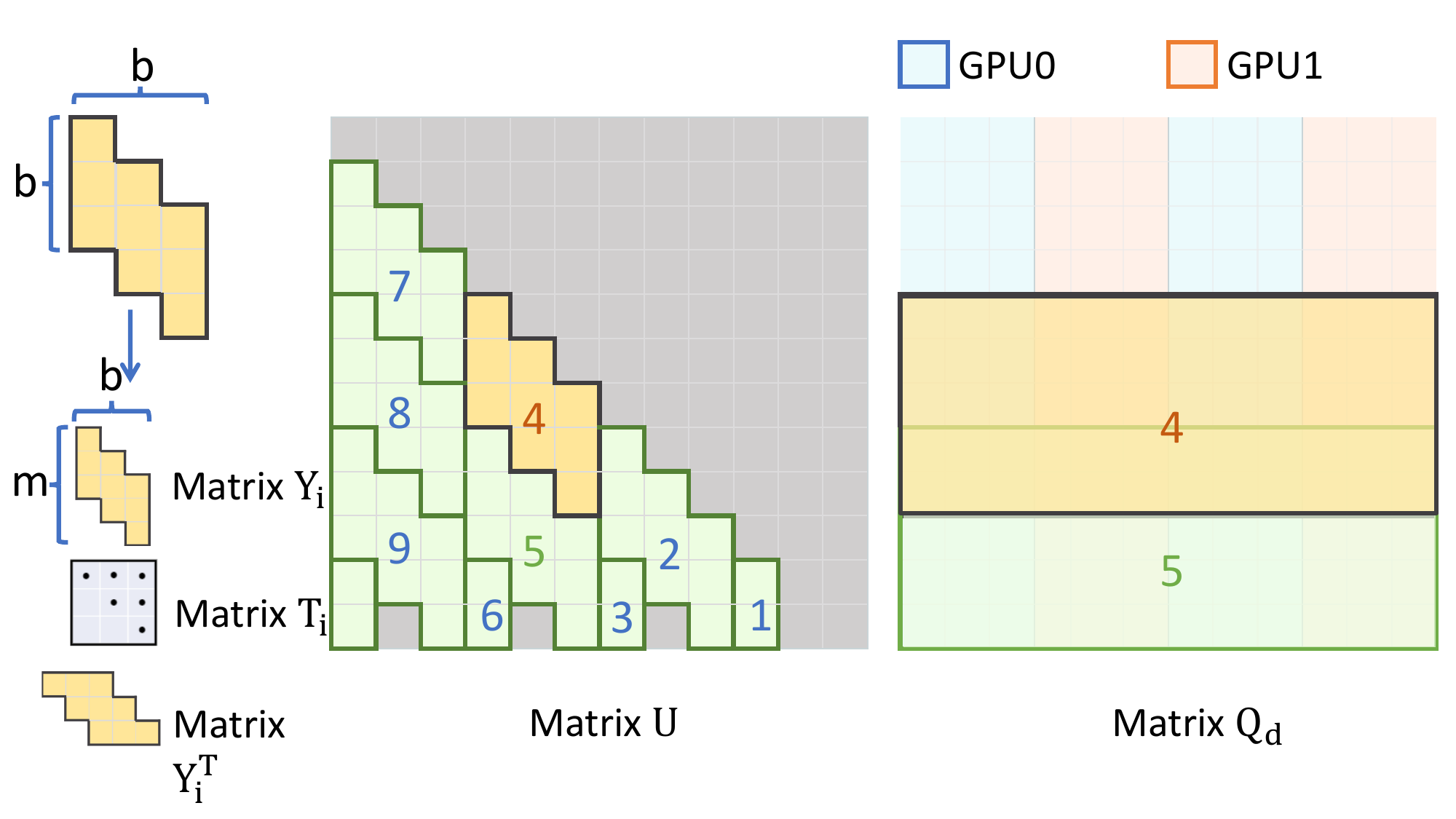}
    \caption{The illustration of BC-Back transformation multiplies $Q_d$.}
    \label{fig:backtrans_blas3}
\end{figure}

When only the eigenvalues are desired, the EVD stops at the end of the iterative solver; while the eigenvectors are also required, back transformation will be taken.
For two-stage EVD, back transformation comprises SBR-Back and BC-Back. While SBR-Back follows the one-stage EVD approach, BC-Back faces strict data dependencies requiring updates in a specific order ($u_{ij}$ after $u_{(i+1)(j-1)}$ and $u_{(i+1)j}$), with a complexity of approximately $2n^3$ FLOPS (see Figure~\ref{fig:backtrans_blas2} for detail). The standard BC-Back implementation lacks BLAS3 operations, limiting performance. MAGMA~\cite{2stage2} optimizes this by grouping Householder vectors via LAPACK's \texttt{larft} routine and performing GEMM-based updates (Figure~\ref{fig:backtrans_blas3}), but BC-Back remains the primary bottleneck, reducing the performance advantage of two-stage EVD over one-stage EVD from 6.1$\times$ to 1.2$\times$~\cite{DBBR}.

\subsection{EVD Solvers on Multi-GPU Architectures}
Multiple linear algebra libraries support computing large scale EVD problems, including ScaLAPACK~\cite{blackford1997scalapack}, Eigen~\cite{EigenLibrary}, Intel MKL on multi-CPUs, and cuSOLVERMp, MAGMA~\cite{MAGMA}, SLATE~\cite{gates2020slate} and ELPA on multi-GPU or hybrid architectures. Basically, on multi-GPU architectures, the matrix are distributed on different GPUs with a 1D block-cyclic style or a tiling style to improve the load balance. Specifically, in terms of the two-stage EVD implemented in MAGMA, SLATE and ELPA, the EVD solvers compute SBR, BC, D\&C, BC-Back and SBR-back in sequence.

However, based on our experiments, all of the EVD solvers on multi-GPU architectures cannot provide good performance and scalability. For example, MAGMA only reaches less than 2\% TFLOPs of the theoretical peak multi-GPUs performance, and using 2 or 4 GPUs even leads to worse performance than single GPU due to the expensive communication overhead. Similar phenomenon is also observed in cuSOLVERMp and SLATE. ELPA is an exception that it keeps good weak and strong scalability, but its performance on 4 GPUs is lower than MAGMA or cuSOLVERMp on one GPU.

\section{Motivation, Difficulties and Solutions in Pipeline}
\subsection{Motivation}

\begin{figure}
    \centering
    \includegraphics[width=1.0\linewidth]{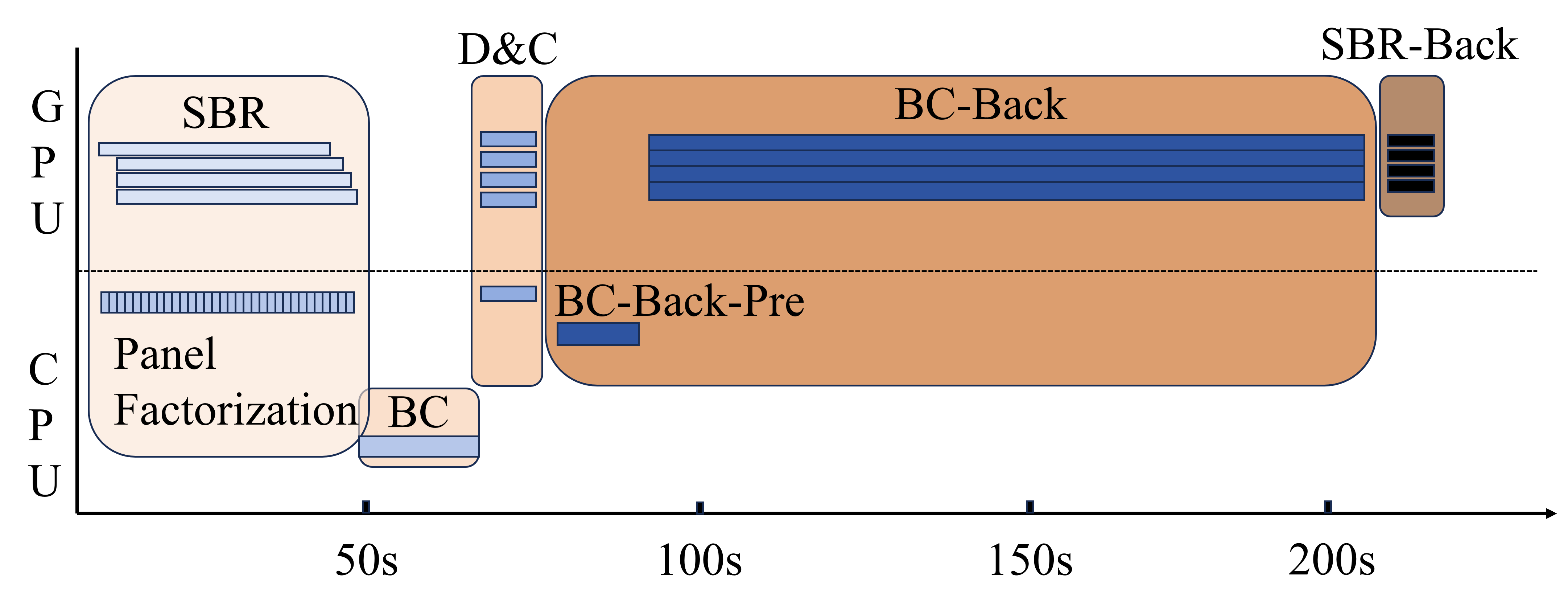}
    \caption{The timeline of MAGMA two-stage EVD routine with matrix size $49152\times 49152$ on 4 A100 GPUs.}
    \label{fig:magma_timeline}
\end{figure}

Figure~\ref{fig:magma_timeline} presents the breakdown of the two-stage EVD on 4 A100 GPUs. MAGMA treats the stages of EVD as independent tasks: for example, BC starts after SBR, and BC-Back is executed at the end of D\&C. This design leads to two issues. First, when performing BC and BC-Back-Pre (forming $T$ matrices with the LAPACK \texttt{larft} routine), the GPUs are idle. Second, SBR suffers from load imbalance, where some GPUs wait for others to finish their SBR tasks. Migrating BC and BC-Back-Pre from the CPU to the GPUs using recent techniques~\cite{DBBR} can mitigate the first issue, however it introduces severe load imbalance (see Section~\ref{sec:dist_bc}). Observing opportunities to overlap computations across stages, pipelining is effective for two reasons: 1) it has the potential to maintain high GPU utilization, and 2) it enables load balancing across the entire EVD pipeline rather than within a single stage, providing more flexibility to adjust per-GPU load. Converting the conventional sequential EVD into a pipelined EVD, however, poses several challenges, which we address in the following sections.

\subsection{Blockwise Distribution}
\begin{figure}
    \centering
    \includegraphics[width=1.0\linewidth]{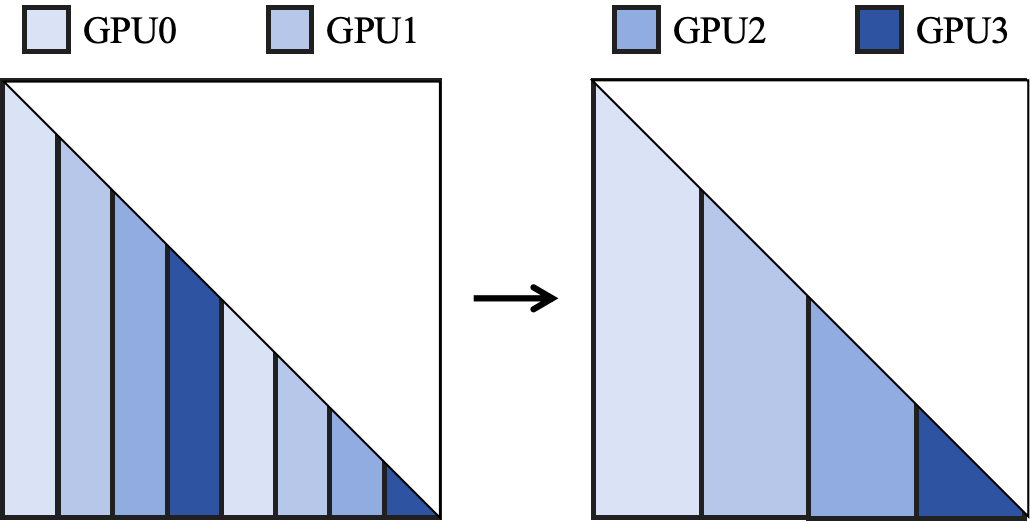}
    \caption{The difference between the block cyclic distribution and the blockwise distribution.}
    \label{fig:distribution}
\end{figure}

Directly converting conventional EVD solvers into a pipelined EVD solver is not easy, because conventional EVD solvers typically adopt a block-cyclic scheme; for example, MAGMA uses a 1D block-cyclic scheme, and cuSOLVERMp employs a 2D block-cyclic scheme. We use SBR as an example to illustrate the difficulty. Figure~\ref{fig:distribution} provides an illustration of MAGMA’s implementation. We observe that GPU0 holds different blocks of $A$, and it will always be busy until the final block on GPU0 is computed. However, the BC process can start as soon as the first few blocks have been updated and computed, and recent fast GPU-based BC implementations require the full GPU resources, especially memory bandwidth~\cite{DBBR}. Although BC can start early, there are still bottlenecks in the block-cyclic scheme: 1) BC cannot fully utilize GPU resources; 2) expensive communication is required to move the band-form matrix between different GPUs; and 3) it is hard to schedule tasks across multiple stages.

Therefore, we abandon the conventional block-cyclic distribution and distribute the matrix using a column-blockwise scheme, as shown on the right of Figure~\ref{fig:distribution}. Compared with the block-cyclic distribution, the blockwise distribution allows us to create a pipeline. For instance, GPU0–GPU3 are busy until the first block in SBR finishes, because all of the GPUs participate in the computations of forming the matrix $Z$ and updating the trailing matrix, which are the most expensive parts of SBR. Once the first block on GPU0 is done, GPU0 can start the following BC stage immediately. With this matrix distribution strategy and the corresponding pipeline, overlap between different stages becomes possible, leading to better utilization of GPU resources.

\subsection{Reordered Stages}

\begin{figure}

    \begin{subfigure}[b]{1.0\columnwidth}
    \centering
    \includegraphics[width=1.0\columnwidth]{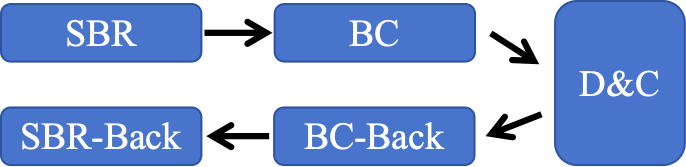}
    \caption{The conventional order in two-stage EVD.\label{fig:conv_order}}
    \end{subfigure}

    \begin{subfigure}[b]{1.0\columnwidth}
    \centering
    \includegraphics[width=1.0\columnwidth]{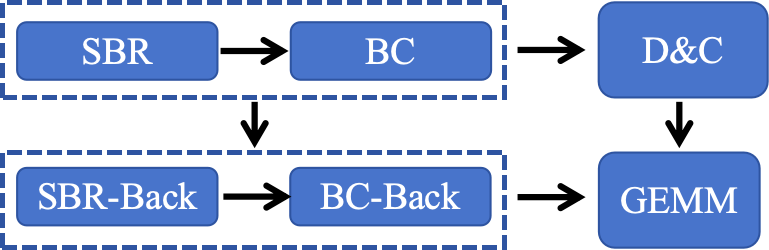}
   
    \caption{The new order in two-stage EVD.\label{fig:reorder}}
    \end{subfigure}

    \caption{The comparison between conventional order and the proposed order in two-stage EVD.\label{fig:RBT}}

\end{figure}

Another difficulty of the pipelined EVD solver is that the D\&C process is hard to be assembled into the pipeline. According to Figure~\ref{fig:backtrans_blas3}, we can find that BC-Back of the first block requires the last several rows in the orthogonal matrix $Q_d$ generated by the D\&C process. Unfortunately, $Q_d$ cannot be obtained before the finalization of D\&C, because $Q_d$ continues to be updated until the last conquer process,  resulting in the stop of the pipeline. To avoid such dependency, we consider to schedule the order of the stages in a different way, based on a simple observation that the D\&C process is actually an independent process to other stages. 

Assuming BC is completed and the tridiagonal matrix is obtained, in the conventional order (Figure~\ref{fig:conv_order}), D\&C will be computed at first to generate the $Q_d$ matrix, and BC-back applies Householder vectors to $Q_d$ to $Q_{bd}$, and SBR-back further applies the WY representation on $Q_{bd}$ to get the final orthogonal matrix $Q$. Indeed, if we regard the D\&C as an independent stage that it can start as soon as the tridiagonal matrix is obtained, then we can bypass the D\&C process and directly move to the SBR-Back and BC-back process to maintain the pipeline, as illustrated in Figure~\ref{fig:reorder}. 
However, as $Q_d$ is not available during the back transformation process, the Householder vectors will be applied on an identity matrix instead of $Q_d$, resulting in an extra GEMM to form the final eigenvectors. Fortunately, the GEMM is square and large, which is efficient on modern GPU architectures, so that the extra overhead is tolerable. Furthermore, as the CPU is almost always idle, we can let the CPU handle D\&C, while the back transformation is executed on GPUs, so that the D\&C can be overlapped ideally. 

\subsubsection{Correctness of Reordered Stages}
One of the problems of reordered stages is that it is unknown whether this modification can still provide correct result. Thus, we try to prove that both the orthogonality and backward error are still bounded by $O(\epsilon)$, where $\epsilon$ is the rounding-off error. For orthogonality, as the Householder operations guarantee $Q_{\text{sb}}= Q_sQ_b$'s orthogonality and $Q_d$ preserves orthogonality via its traditional D\&C computation, we only need to prove the orthogonality of the GEMM operation $Q = Q_{\text{sb}} Q_d$.
Let \(E=\text{fl}(Q_1Q_2)-Q_1Q_2\). Then:
$$
\text{fl}(Q_1Q_2)\text{fl}(Q_1Q_2)^T-I=(Q+E)(Q+E)^T-I=EQ^T+QE^T+EE^T.
$$
Therefore:
$$\|\text{fl}(Q_1Q_2)\text{fl}(Q_1Q_2)^T-I\|_2=\|EQ^T+QE^T+EE^T\|_2{\leq}2\|EQ^T\|_2+\|E\|_2^2.$$
From~\cite{MatrixComputation}, we have:
$$\|\text{fl}(QA)-QA\|_F\leq\sqrt{n}\epsilon\|A\|_F.$$
When \(A\) is orthogonal, \(\|A\|_F=\sqrt{n}\), and\(\|A\|_2\leq\|A\|_F\), so:$$\|E\|_2\leq\|E\|_F{\leq}n\epsilon.$$
Since orthogonal transformations preserve the 2-norm, we have\(\|EQ^T\|_2=\|E\|_2\).
Therefore:$$\|\text{fl}(Q_1Q_2)\text{fl}(Q_1Q_2)^T-I\|_2\leq2n\epsilon+n^2\epsilon^2.$$

The relative orthogonality error:$$\frac{\|\text{fl}(Q_1Q_2)\text{fl}(Q_1Q_2)^T-I\|_2}{n}\leq2\epsilon.$$

This shows the GEMM \(Q_1Q_2\) remains orthogonality. 
For backward stability: $$\text{fl}(Q_1Q_2)\Sigma\text{fl}(Q_1Q_2)^T-Q{\Sigma}Q^T=E{\Sigma}E^T+E{\Sigma}Q^T+Q{\Sigma}E^T.$$

where \(Q =Q_1Q_2\), \(E=\text{fl}(Q_1Q_2)-Q\). Then:$$\|\text{fl}(Q_1Q_2)\Sigma\text{fl}(Q_1Q_2)^T-Q{\Sigma}Q^T\|_2\leq\|E{\Sigma}E^T\|_2+2\|{\Sigma}E\|_2.$$

Since \(\Sigma\) is diagonal, \(\|E\|_2{\leq}n\epsilon\), we derive:$$\|\text{fl}(Q_1Q_2)\Sigma\text{fl}(Q_1Q_2)^T-Q{\Sigma}Q^T\|_2\leq(2n\epsilon+n^2\epsilon^2)\|\Sigma\|_2.$$

Therefore:$$\frac{\|\text{fl}(Q_1Q_2)\Sigma\text{fl}(Q_1Q_2)^T-Q{\Sigma}Q^T\|_2}{n\|Q{\Sigma}Q^T\|_2}\leq2\epsilon+n\epsilon^2\approx2\epsilon,$$

indicating the backward stability is preserved.

Experimentally, Table~\ref{tbl:accu_comp2} compares the backward error $\frac{||A-Q\Lambda Q^T||}{n||A||}$ and orthogonality $\frac{||I-QQ^T||}{n}$ between reordered stages EVD and cuSOLVER \texttt{Dsyevd} routine with matrix size $16384\times 16384$. And the results reveal that conventional EVD has slightly better stability and orthogonality than reordered stages EVD, but reordered stage EVD still preserves the FP64 accuracy, which conforms the above mathematical proof.

\begin{table}[h]
    \centering
    \begin{tabular}{|c|c|c|c|c|}
        \hline
        \multirow{2}{*}{Distribution} & 
        \multicolumn{2}{c|}{Backward} & 
        \multicolumn{2}{c|}{Orthogonality} \\ \cline{2-5} 
        & cuSOLVER & Ours & cuSOLVER & Ours \\ \hline
        Cluster0     & 5.5E-19& 1.4E-18
& 8.2E-17& 1.5E-16\\ \hline
        Cluster1     & 4.4E-19& 3.9E-19
& 7.9E-17& 1.4E-16\\ \hline
        Geometric    & 2.5E-19& 8.6E-19
& 6.7E-17& 1.4E-16\\ \hline
        Arithmetic   & 5.6E-19& 1.3E-18
& 8.5E-17& 1.5E-16\\ \hline
        Normal       & 3.8E-19& 6.7E-19
& 4.0E-17& 1.5E-16\\ \hline
        Uniform      & 4.9E-19& 1.8E-18& 4.0E-17& 1.5E-16\\ \hline
    \end{tabular}
   \caption{The EVD accuracy comparison between cuSOLVER and the proposed reordered stages EVD with different matrix types. For Cluster0, Cluster1, Geometric and Arithmetic distribution, condition number is 1E8 and the largest eigenvalue is 1E6.}\label{tbl:accu_comp2}
\end{table}

\subsection{Load Balance and Overall Pipeline}

However, there is also an issue existing in the proposed method, that it has severe load imbalance. This is because GPU0 only needs handle the Block0, while GPU1 not only handles the trailing matrix update in Block0, but also needs to handle its own computations. Additionally, the BC process also has imbalance load. Thus, regarding tridiagonalization, GPU$_i$ always undertakes more computations than GPU$_{i-1}$.

\begin{figure}
    \centering
    \includegraphics[width=1\linewidth]{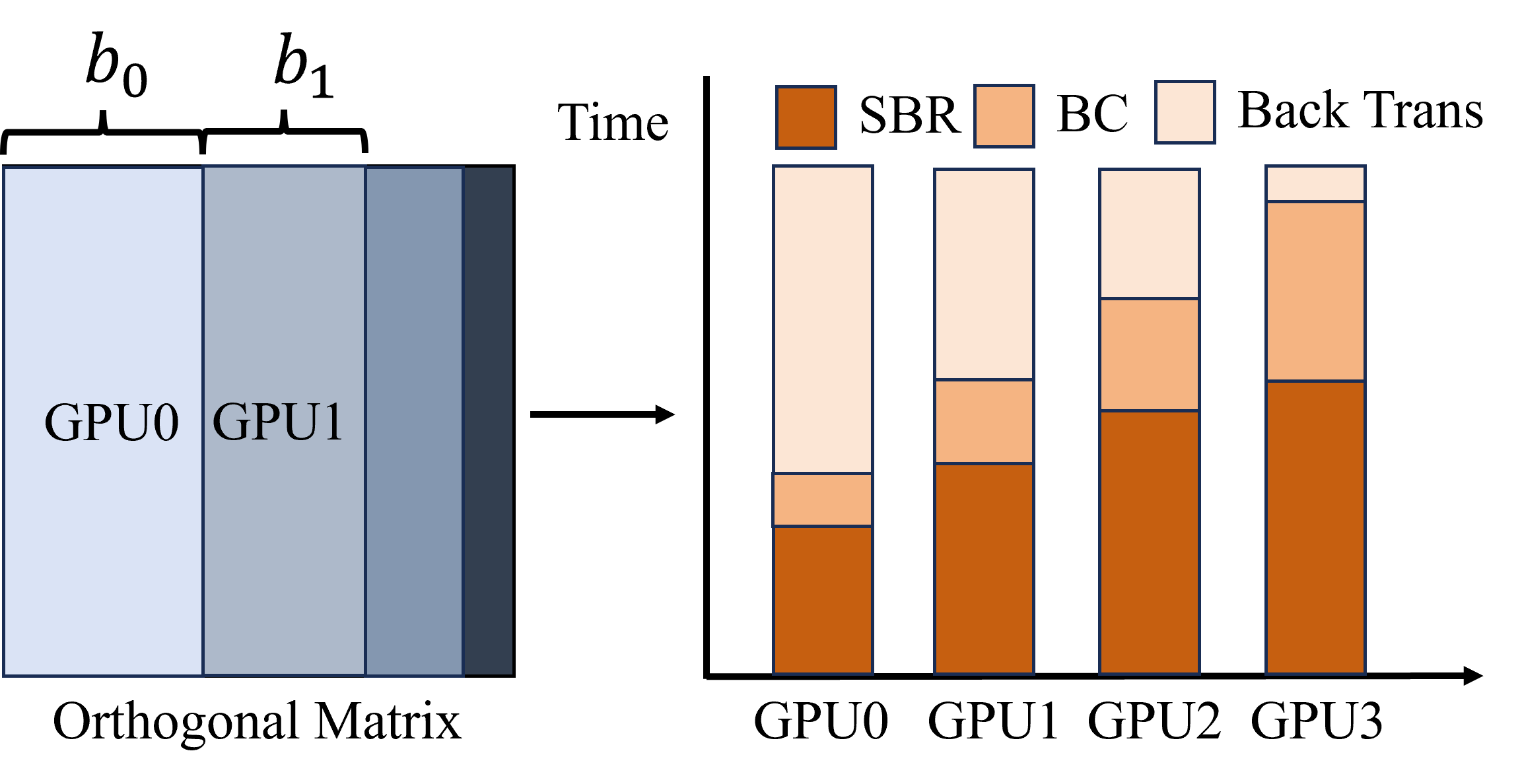}
    \caption{The illustration of adjustable blocksize in SBR-Back and BC-Back.}
    \label{fig:loading_balance}
\end{figure}

Fortunately, benefiting from the reordered-stage strategy which detaches the D\&C process, we have an opportunity to leverage SBR-Back and BC-back to balance the GPU load, and the idea shown in Figure~\ref{fig:loading_balance} is natural and simple. Compared to SBR and BC, the back transformation process including SBR-Back and BC-Back has less dependency, which means we have more flexibility to form the eigenvectors. As Figure~\ref{fig:loading_balance} shows, suppose there are 4 GPUs, we can dynamically choose $b_0>b_1>b_2>b_3$, so that the GPU$_i$ performs more computations than other GPUs than GPU$_{i+1}$. Hence, demonstrated by the bar figure in Figure~\ref{fig:loading_balance}, although GPU0 spends less time on SBR and BC, we can let GPU0 do more computations in back transformation, leading to end-to-end load balance. In practice, we keep the per-GPU adjustment within 5\% of a base block size to provide enough slack to equalize the timeline.

\begin{figure}
    \centering
    \includegraphics[width=1\linewidth]{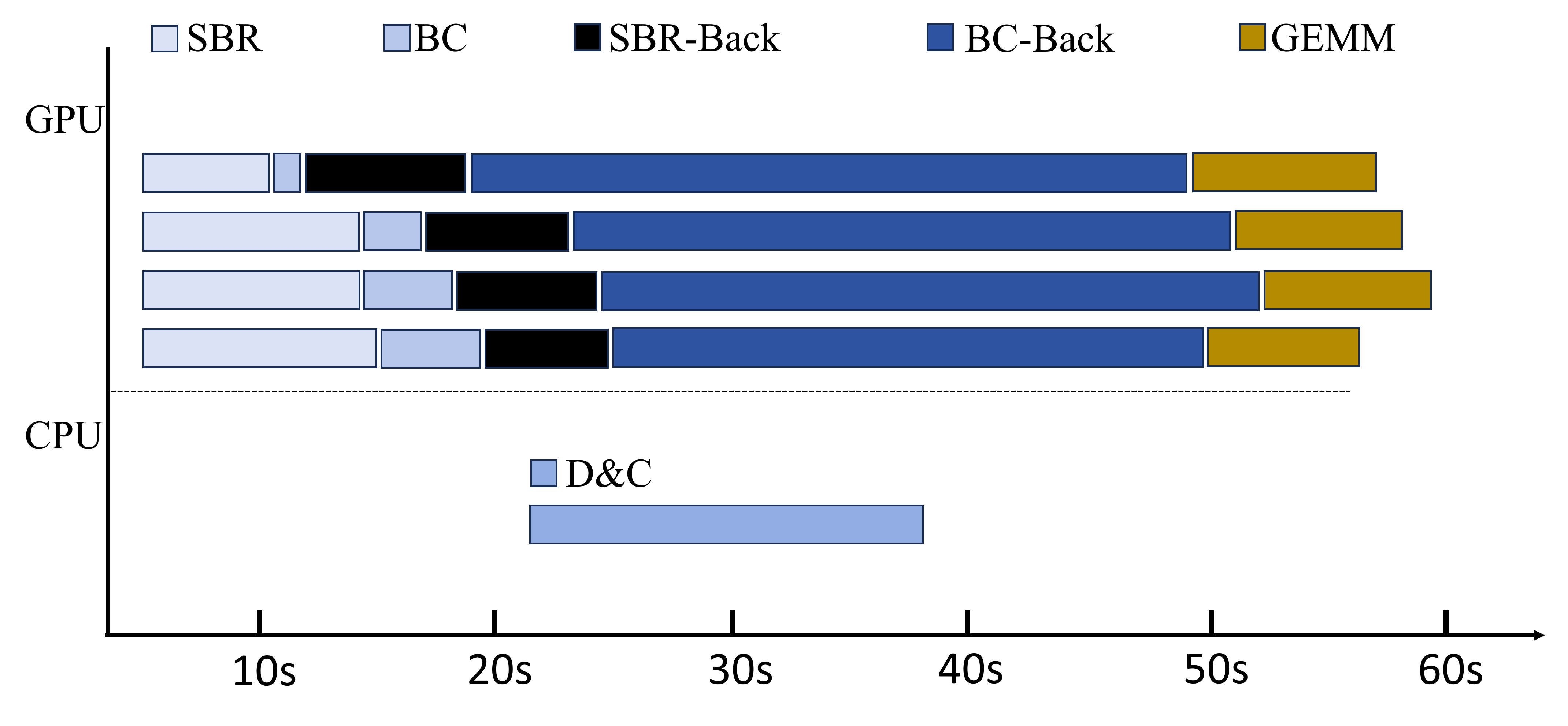}
    \caption{The overall pipeline of the proposed EVD solver.}
    \label{fig:new_pipeline}
\end{figure}

After adjusting the blocksize in back transformation process, the rough overall pipeline is shown in Figure~\ref{fig:new_pipeline}.  Compared to MAGMA, we do not pursue the load balance in a single stage, instead, we try to leverage the back transformation to balance the load with tridiagonalization. As a result, although the tridiagonalization process has severe load imbalance, the overall load can be balanced.

Finally, the initial pipelined two-stage EVD algorithm (Algorithm~\ref{alg:alg1_naive}) employs MPI with one process per GPU (identified by $rankID$). Each GPU retrieves its local $A$ submatrix (Lines 5–6). The root process (rank 0) executes SBR immediately (Lines 8–14), while others wait for preceding ranks. Crucially, after the completion of SBR, the GPU immediately proceeds to BC without waiting for the final finish of SBR, and then concurrently accumulating $W$ for SBR-Back (Lines 15–16). Since BC (it is inherently fast ) always completes before SBR-Back, BC-Back can initiate immediately. Leveraging CPU idle time during SBR-Back and BC-Back, rank 0 spawns a D\&C thread (Lines 29–32) for parallel execution. And after D\&C and BC-Back's finalization, the final GEMM will assemble the global eigenvector matrix $Q$.

\begin{algorithm}
\caption{The Pipelined Multi-GPU EVD\label{alg:alg1_naive}}
\begin{algorithmic}[1]
\REQUIRE 
    Matrix $A$ -- the original dense symmetric matrix. \\
    Int $n$ -- the row or column numbers of matrix $A$. \\
    Int $n\-gpu$ -- the number of GPUs. 
 
\ENSURE 
    Array $\Lambda$ -- the eigenvalues of matrix $A$. \\
    Matrix $Q$ -- the orthogonal eigenvector matrix of $A$.

\STATE \textcolor{blue}{\% Initialize the MPI and get the process rank.}
\STATE $rankID \gets MPI\_Init()$
\STATE $n_{cols} \gets \frac{n}{n_{gpu}}$

\STATE \textcolor{blue}{\% Partition the data using block-wise style and transfer each block to GPU.}
\STATE $A_{ih} = A(:, rankID*n_{cols} : (rankID+1)*n_{cols})$
\STATE $A_{id} = cudaMemCopy(A_{ih})$

\STATE \textcolor{blue}{\% SBR subprocess}
\STATE if $0 == rankID$
\STATE \hspace{\algorithmicindent} $SBR(A_{id})$
\STATE else
\STATE \hspace{\algorithmicindent} \textcolor{blue}{\% wait the finish of $rankID-1$'s SBR.}
\STATE \hspace{\algorithmicindent} $waitSBRFinishedEvent(rankID - 1)$
\STATE \hspace{\algorithmicindent} $SBR(A_{id})$
\STATE {endif}

\STATE $T, U_i = BC(A_{id})$ \textcolor{blue}{\% BC subprocess}

\STATE $W_i =  SBR\_Back\_genW(W_{id})$ \textcolor{blue}{\% SBR\_Back\_genW subprocess}

\STATE \textcolor{blue}{\% BC subprocess}
\FOR{$k = 1:1:n_{gpu}$}
\STATE $Q_{is}^T=SBR\_Back\_genQ()$

\STATE $waitSBR\_Back\_genWFinishedEvent(k+1)$
\ENDFOR

\STATE \textcolor{blue}{\% wait for the finish of the last GPU's BC.}
\STATE $waitBCFinishedEvent(n_{gpus})$

\STATE \textcolor{blue}{\% Collect the U matrices generated by BC on each GPU.}
\STATE $U = MPI\_gather(U_i)$

\STATE \textcolor{blue}{\% BC-Back subprocess}
\STATE $Q_{ibs}^T = BC(U)$

\STATE \textcolor{blue}{\% The rankID 0 spawns a thread for D\&C, enabling concurrent execution with back transformation.}
\STATE if $0 == rankID$
\STATE \hspace{\algorithmicindent} $Z, \Lambda  = std::thread DC\_thread(CPU_DC(T)$
\STATE \hspace{\algorithmicindent} $std::thread.join()$
\STATE {endif}

\STATE \textcolor{blue}{\% Synchronize all Ranks.}
\STATE $MPI\_barrier();$

\STATE \textcolor{blue}{\% Final GEMM}
\FOR{$k = 1:1:n_{gpu}$}
\STATE $Z_i = cudaMemCopyAsync(Z(:, (k-1)*n_{cols}:k*n_{cols}))$

\STATE $Q_i(k*n_{cols}:(k+1)*n_{cols}),:) = cudaGEMM(Z_i, Q_{ibs})$
\ENDFOR

\STATE \textcolor{blue}{\% Assemble the global eigenvector matrix $Q$.}
\STATE $Q(rankID*n_{cols}:(rankID+1)*n_{cols}), :) = Q_i$

\STATE $MPI\_Finalize$

\end{algorithmic}
\end{algorithm}
\section{Implementation and Optimization}
Although the pipelined design of EVD provides a better utilization of GPU resources, the implementations and optimizations are also important to further improve the performance. Therefore, in this section, we'll discuss our implementation and optimization details on some stages.
\subsection{Communication Avoiding SBR}

\begin{figure}
    \centering
    \includegraphics[width=1\linewidth]{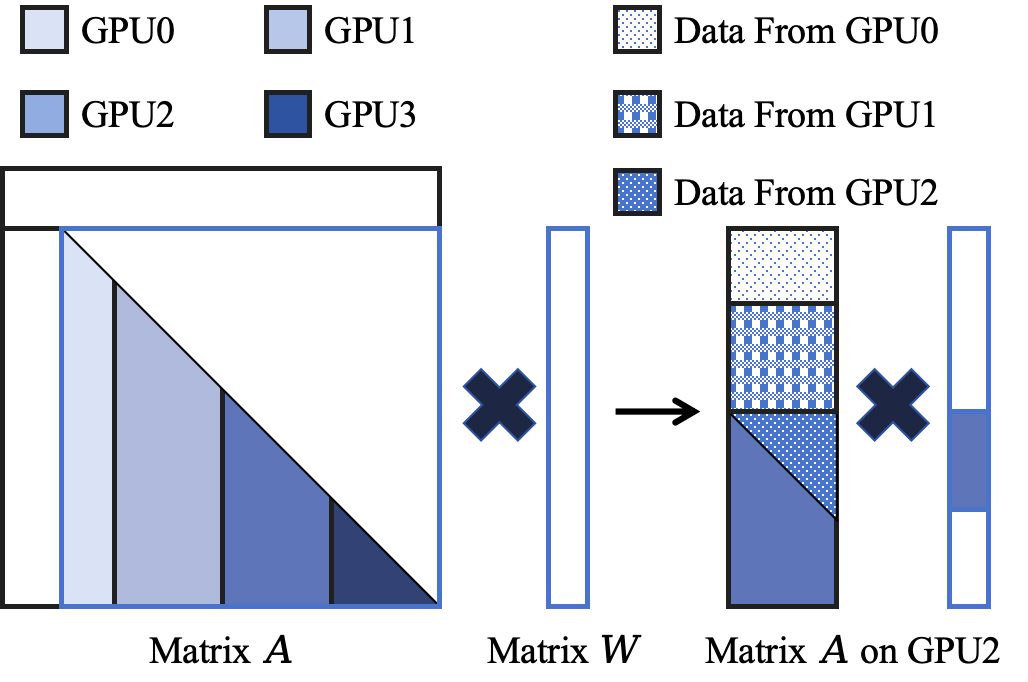}
    \caption{The communication behavior when only storing the lower triangular part of matrix $A$.}
    \label{fig:SBR_comm}
\end{figure}

Compared to SBR on a single GPU, SBR on multi-GPUs is more challenging due to communication on the trailing matrix. Considering the computations of forming $Z=AW-\frac{1}{2}YW^TAW$ and based on our previous design, $A$ is distributed on GPUs blockwisely, and $W$ is already broadcast to all GPUs. Then computing $AW$ is challenging if only the lower triangular part of $A$ is stored, because expensive communication always exist if GPU$_i$ doesn’t hold the corresponding rows in the upper triangular part of $A$. Figure~\ref{fig:SBR_comm} gives an explanation, suppose GPU2 undertakes a sub-task, to deliver the correct results, GPU2 needs the data from GPU0 and GPU1 to fill in the upper part of the local matrix, leading to communication. Furthermore, this kind of communication happen very frequently because $AW$ will be recomputed after every panel factorization. Quantitatively, if the bandwidth is $b$ and the matrix size is $n\times n$, then the total communication words will be $\sum_{i=1}^{\frac{n}{b}-1} \frac{1}{2}(n-i*b)^2=\frac{n(n-b)(2n-b)}{12b}$, which is intolerable.

To solve this problem, one can consider to distribute the entire matrix rather than a triangular matrix, so that the local matrix on each GPU can hold the entire matrix. The disadvantage is that the trailing matrix updates using \texttt{syr2k} will be substituted with GEMM, thereby the FLOPS are increasing to 2x, but we demonstrate the scarification is worthy. We can specify GPU$_i$ as an example. If the matrix is stored as a full matrix, then the extra computation cost of trailing matrix update will be $t_1=\frac{2(ik-xb)kb}{p}$, where $k$ is the number of columns each GPU takes, $b$ is the bandwidth in BC, $x$ is the round of panel factorization in SBR and $p$ is the TFLOPs of the GEMMs. Then the communication time will be $t_2=\frac{8(ik-xb)*k}{q}$, where and $q$ is the network bandwidth. If $t_1<t_2\rightarrow b<\frac{4p}{q}$,  then the extra time cost of computations will be less than the communication using triangular form. On the modern GPU architectures, computing $AW$ typically can reach 10 TFLOPs, while $q\approx 0.35~TB/s$ on H100-SXM GPU with Gen4 Nvlink one direction. Thus, if $b<114$, the proposed SBR spends less time. Typically, in two-stage EVD, $b$ is usually set to 32~\cite{DBBR} to deliver best performance, which indicates performing extra computations in SBR is worthy, because the inter-GPU communication are more expensive. Note that using GPUs without Nvlink leads to smaller $q$, thereby amplifying the advantage of the proposed SBR.

After this optimization, the trailing matrix $A$ doesn’t need any communication, while only the matrices $W, Y$ and $Z$ are expected to broadcast, which accounts for $\sum_{i=1}^{\frac{n}{b}-1} 3(n-i*b)*b$. To further improve the performance, the GEMM is upgraded to \texttt{syr2k} on the last GPU, as the last GPU doesn’t need data from other GPUs anymore. Also, we adopt double blocking SBR~\cite{DBBR} to accelerate the trailing matrix update process.

\subsection{Distributed BC\label{sec:dist_bc}}
Compared to distributed SBR, BC is a more difficult task to be distributed. As a result, even the recent BC implementations on distributed computer architectures including ELPA~\cite{marek2014elpa} and MAGMA~\cite{MAGMA} still handle the BC problem on a single CPU, which is slow and a waste of hardware resource. Wang et.al.~\cite{DBBR} proposes a fast single GPU-based implementation, but we suppose that directly employing the same implementation (each warp handles one BC sweep) incurs large communication overhead and synchronization problems. For example, if we let each GPU handle a group of sweeps, then every GPU will have to load the entire band matrix $B$. Besides, as each Householder transformation in BC will affect the next $2b\times b$ block in $B$, then all of the GPUs have to update this block, thereby resulting in large communication overhead.

\begin{figure}
    \centering
    \includegraphics[width=1\linewidth]{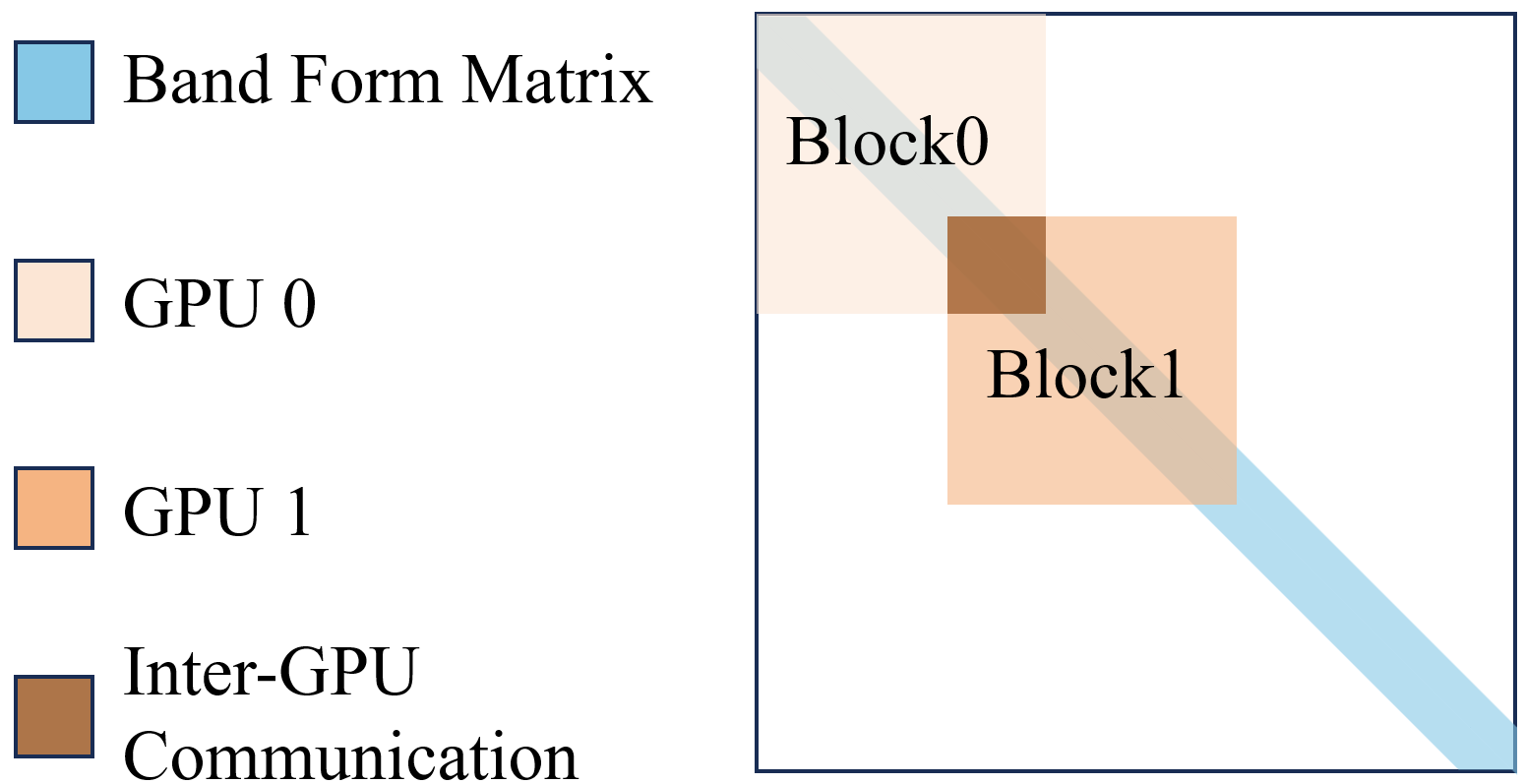}
    \caption{Illustration of distributed BC implementation.}
    \label{fig:distributed_BC}
\end{figure}

Thus, we propose a new BC algorithm on distributed GPU architectures without large communication, and the idea is shown in Figure~\ref{fig:distributed_BC}.  To simplify the illustration, we use two GPUs as an example, and suppose GPU0 and GPU1 takes Block0 and Block1 from band form matrix respectively. GPU0 starts performing BC after the finalization of SBR task on GPU0, while  GPU1 is still working on its own SBR task meanwhile. Once BC on GPU0 reaches the bottom right of Block0 (brown block in Figure~\ref{fig:distributed_BC}), GPU1 can finish its own SBR task and wait for synchronization. As Figure~\ref{fig:distributed_BC} shows, the brown block is an overlap of Block0 and Block1, thereby needing inter-GPU communication to deliver the up-to-date data to GPU1. Fortunately, it is the only communication between subsequent GPUs, while other distribution strategies such as block-cyclic require communication between all GPUs on the entire band matrix.

\subsection{BLAS2-based BC-Back}
The aforementioned BC-Back implemented in MAGMA converts BLAS2 to BLAS3 operations to improve the performance. However, as Figure~\ref{fig:backtrans_blas3} shows, MAGMA's BC-Back relies on LAPACK \texttt{larft} routine to form the $YTY^T$, and we can find that  $T$ is a triangular matrix, and $Y$ is padded with zeros. As a result, extra computations are taken, and the actual FLOPS increase from $2mn^2$ to $4mn^2$. Although BLAS3 has better performance than BLAS2, it is unknown if increasing the FLOPS is worthy. Based on our testing on GEMMs in MAGMA's BC-Back, we find that, due to the special GEMM shapes (e.g. two dimensions are 64 and another dimension is very large ), the GEMMs can only reach up to 6 TFLOPs on H100 GPU, while the peak performance is 67 TFLOPs. Besides, another overhead is typically overlooked that calling LAPACK \texttt{larft} is not cheap and it typically costs 20\% time of BC-Back. In this case, we have an opportunity to design an elaborate BC-Back kernel using BLAS2 operations, it will be potentially faster than MAGMA if the BLAS2 operations can reach 3 TFLOPs.

\begin{figure}
    \centering
    \includegraphics[width=1.0\columnwidth]{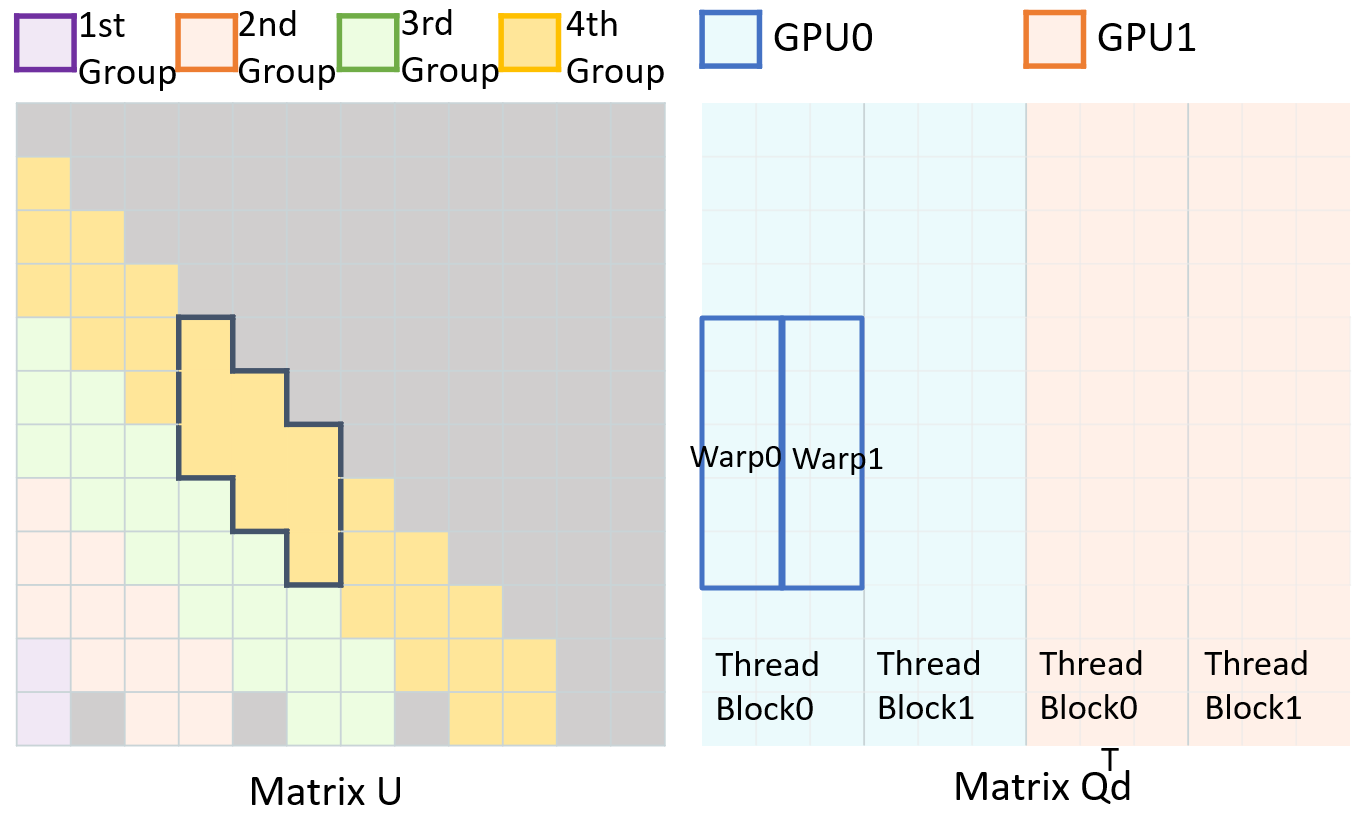}
    \caption{The proposed BC-Back implementation.\label{fig:alg1_detail}}
\end{figure}

Figure~\ref{fig:alg1_detail} details the design of BLAS2-based BC-Back. The matrix $U$ stores the Householder vectors generated from BC, and the matrix $Q_s^T$ represents the orthogonal matrix from SBR-Back. As BC-Back has strict data dependency within $U$ matrix, we follow the bottom to top and left to right order to update $Q_s^T$. To further improve the performance, on a single GPU, we optimize the CUDA kernel carefully using the following techniques.
\begin{itemize}
    \item Collecting several Householder vectors into a group and loading several groups into shared memory together. For example, as shown in Figure~\ref{fig:alg1_detail},  we load $G_1B_1$, $G_2B_1$, $G_3B_1$ and $G_4B_1$ into shared memory at the same time, so that they can be reused until all columns in $Q_s^T$ are updated. 
    \item Loading $Q_s^T$ into registers instead of shared memory. Compared to matrix $U$, $Q_s^T$ is updated frequently and needs to be exchanged between register files and global memory one round after another. Therefore, storing $Q_s^T$ in register files has two distinct advantages: 1) save more shared memory space for storing $U$, and 2) moving data from global memory to registers is faster than that to shared memory on many GPU architectures.
    \item Avoiding bank conflicts in shared memory. The length of Householder vectors in $U$ is typically a multiple of 32, so that it is easy to incur bank conflicts. To solve this problem, we rearrange the elements in $U$.

\end{itemize}

\begin{figure}
    \centering
    \includegraphics[width=1.0\columnwidth]{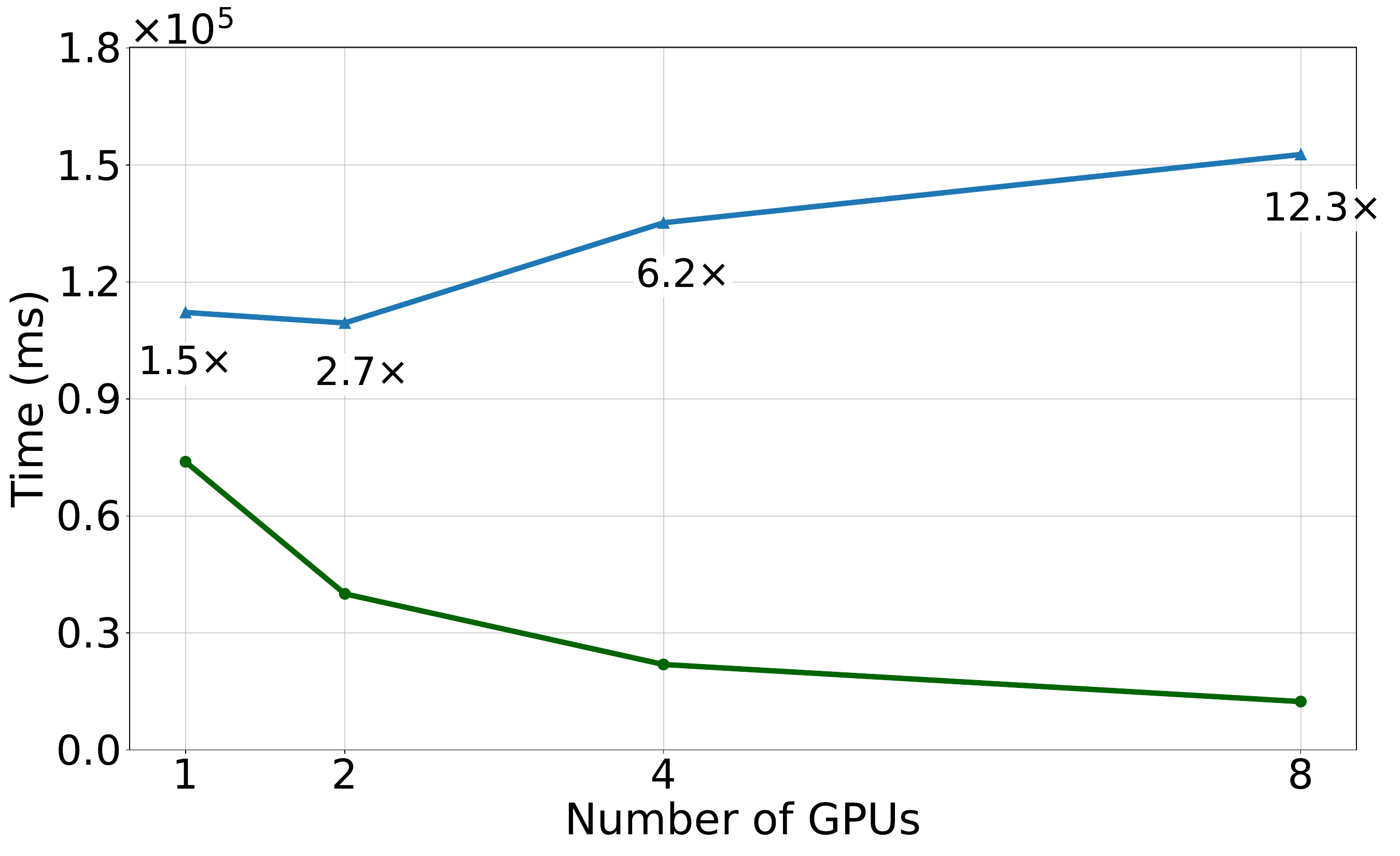}
    \caption{The performance comparison between MAGMA BC-Back and proposed BC-Back on A100 GPUs. Matrix size is $49152\times 49152$.\label{fig:line_bc_back}}
\end{figure}

With the above techniques, our BC-Back kernel provides 1.5x speedup compared to MAGMA's BC-Back on 1 A100 GPU, as shown in Figure~\ref{fig:line_bc_back}. Besides, based on the experiments, we find MAGMA's BC-Back does not have strong scalability, resulting in 12.3x speedup on 8 A100 GPUs. This is probably because the BLAS3 implementation in MAGMA generates $Y$ and $T$ matrices (as depicted in Figure~\ref{fig:backtrans_blas3}), thereby increasing communication and leading to difficulties on hiding the data movement latency. 

\section{Evaluation}
Experiments were conducted on two Multi-GPU architectures: an 8-GPU NVIDIA A100 PCIe platform and an 8-GPU NVIDIA H100-SXM platform. Both platforms utilize identical 48-core Intel\textsuperscript{\textregistered} Xeon\textsuperscript{\textregistered} Platinum 8468 processors and ran Debian GNU/Linux 12 (bookworm). All evaluations employed the NVIDIA HPC SDK 25.3, which integrates a C++ compiler with optimized cuBLAS, cuSOLVER and NVSHMEM libraries.

\subsection{Overall Performance}
\begin{figure}
   \begin{subfigure}[b]{1.0\columnwidth}
    \includegraphics[width=\textwidth]{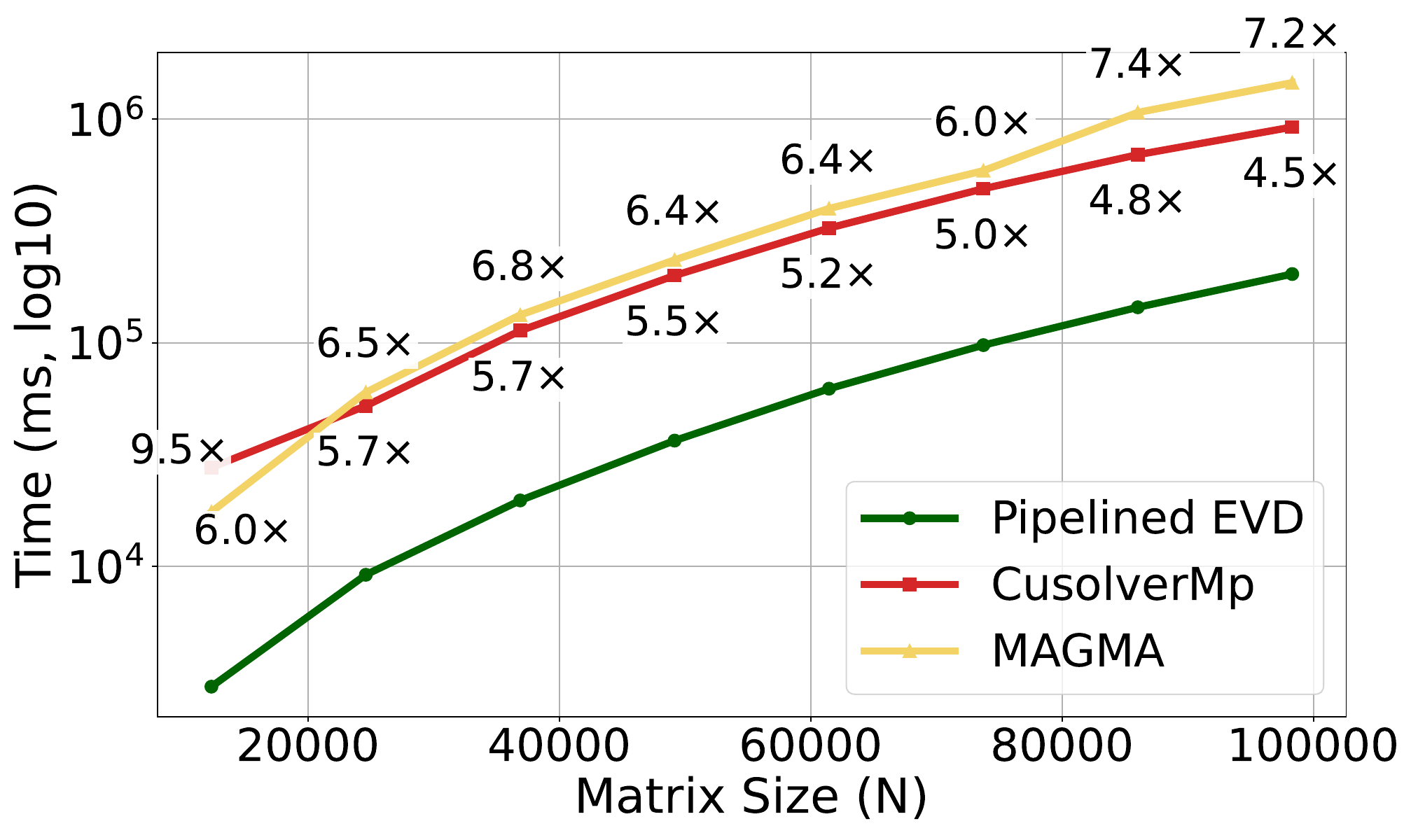}
    \caption{The overall EVD performance on 8 A100 GPU.}
     \label{fig:A100_perf}
    \end{subfigure}
    
      \begin{subfigure}[b]{1.0\columnwidth}
    \includegraphics[width=\textwidth]{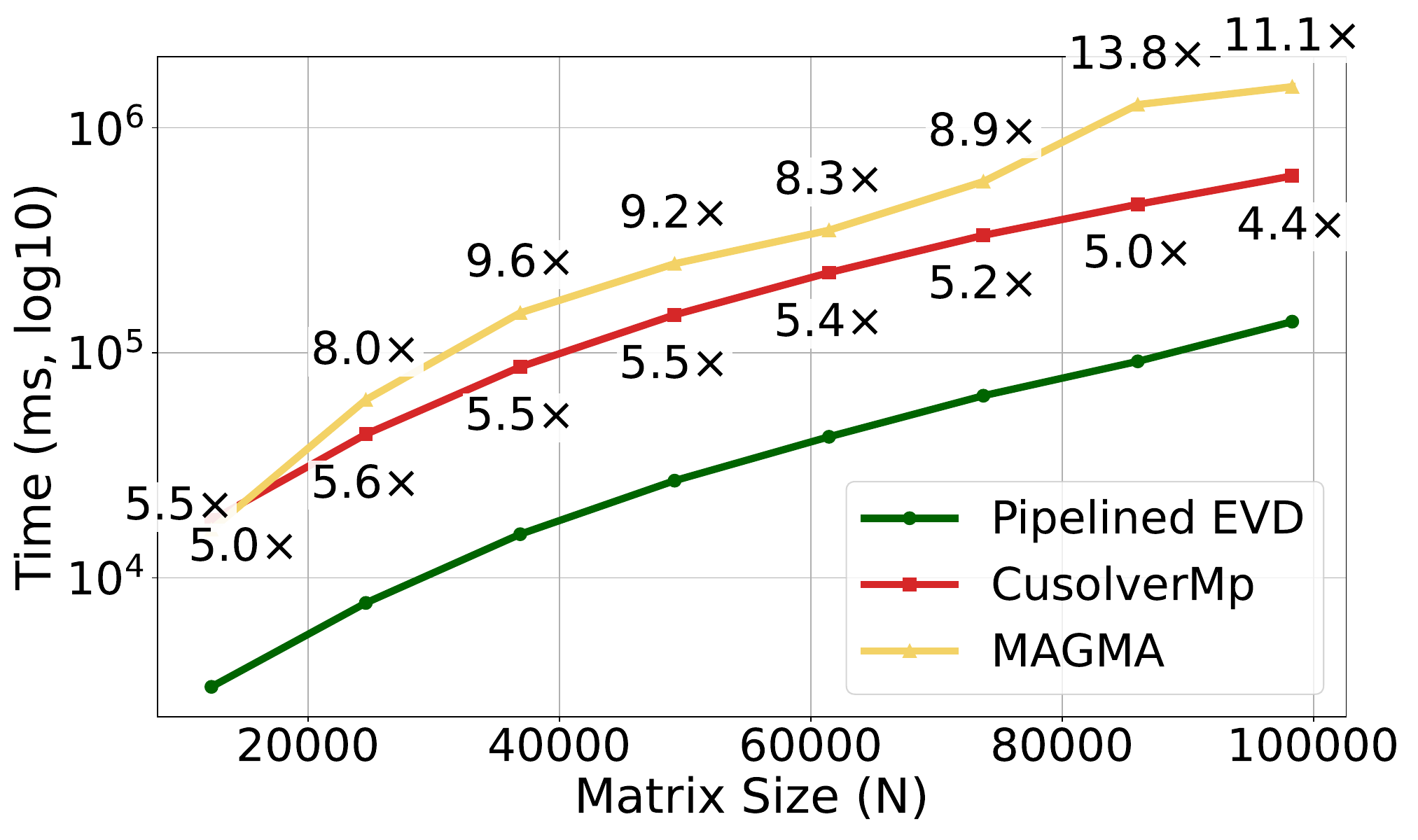}
    \caption{The overall EVD performance on 8 H100 GPU.}
     \label{fig:h100_perf}
\end{subfigure}
\caption{The overall performance comparison between cuSOLVERMp, MAGMA, and pipelined EVD for different matrix sizes on 8 A100/H100 GPUs.}
\label{fig:evd_perf}
\end{figure}

We compare the proposed pipelined EVD algorithm against two implementations, cuSOLVERMp and MAGMA in Figure~\ref{fig:evd_perf}. In short, the pipelined EVD demonstrates consistent and substantial performance gains on A100 and H100-SXM GPUs. On the A100 GPUs, it achieves 5.74$\times$ and 6.59$\times$ speedup over cuSOLVERMp and MAGMA; while on the H100 GPUs, it delivers 5.25$\times$ and 9.24$\times$ speedup relative to cuSOLVERMp and MAGMA, respectively.

These results validate two critical advantages of our approach: first, the algorithmic efficiency derived from the pipelined design, which significantly reduces synchronization overhead while maximizing GPU utilization through an optimized workflow; and second, the hardware generality demonstrated by consistent performance gains across fundamentally different GPU architectures, confirming that its portability is not limited to any specified GPU architecture.

\subsection{Scalability}
\begin{figure}
   \begin{subfigure}[b]{1.0\columnwidth}
    \includegraphics[width=\textwidth]{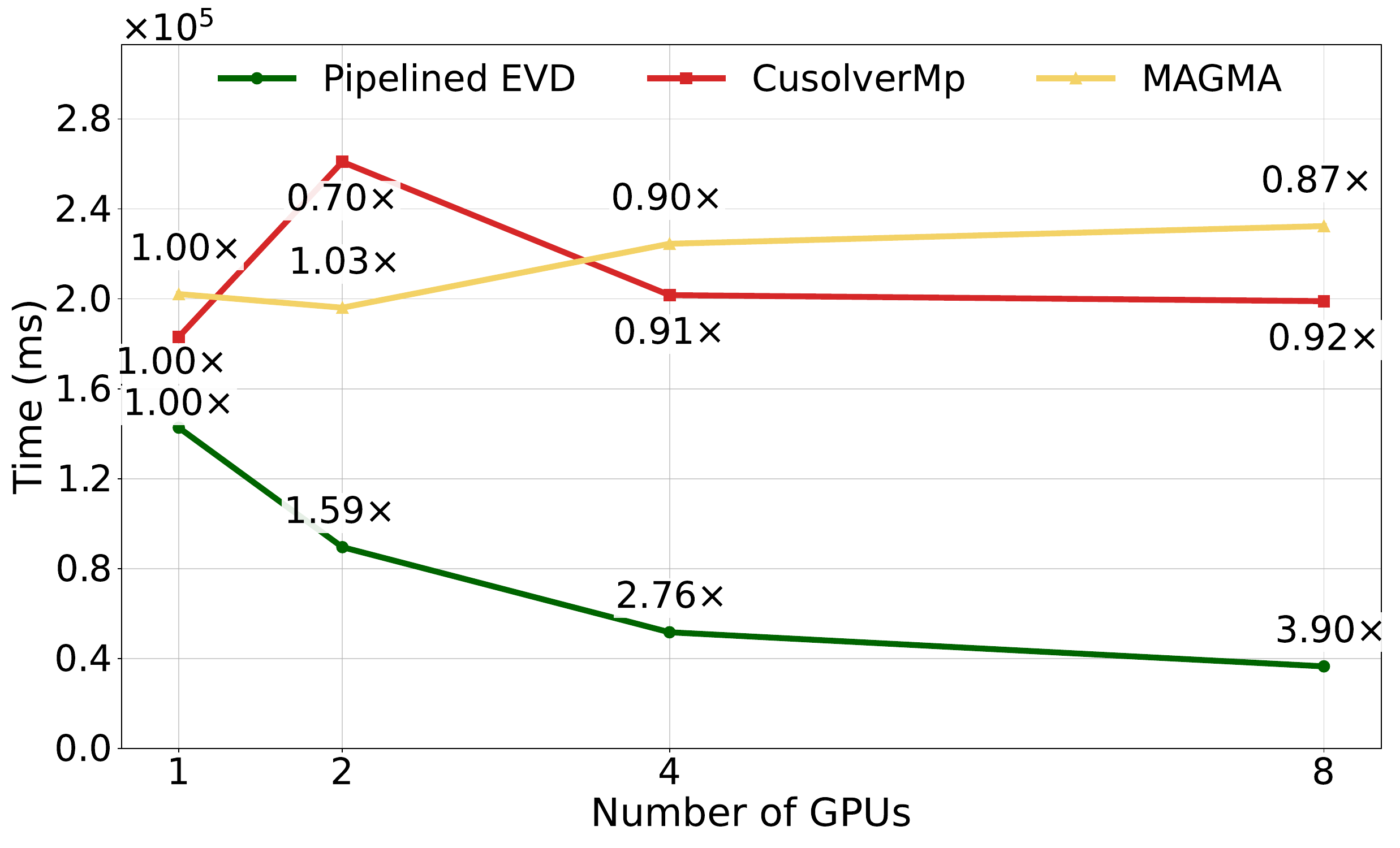}
    \caption{The strong scalability comparison on A100 GPUs.}
     \label{fig:a100_strong}
    \end{subfigure}
    
      \begin{subfigure}[b]{1.0\columnwidth}
    \includegraphics[width=\textwidth]{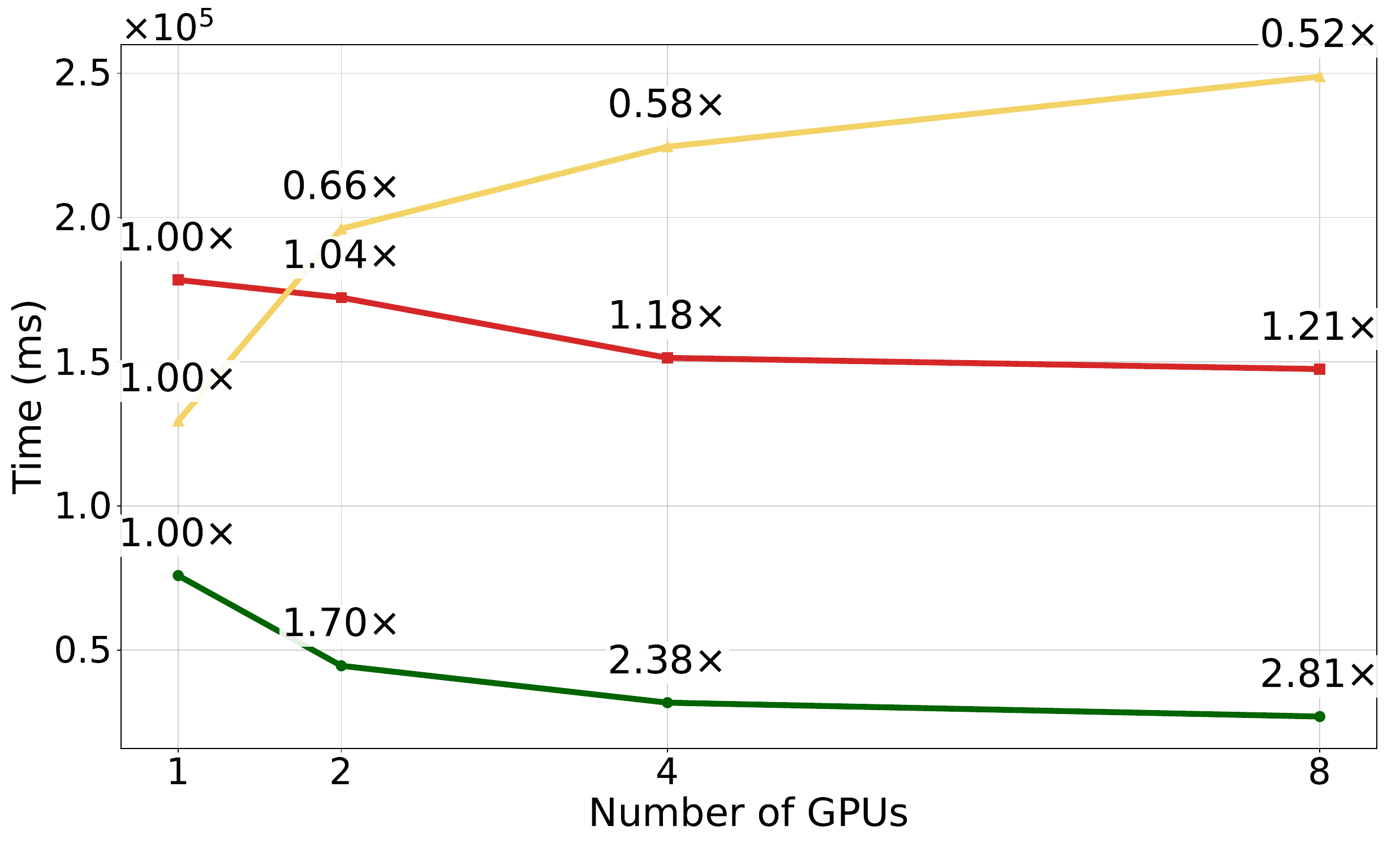}
    \caption{The strong scalability comparison on H100 GPUs.}
     \label{fig:h100_strong}
\end{subfigure}
\caption{The strong scalability comparison among cuSOLVERMp, MAGMA, and the our proposed pipelined EVD.}
\label{fig:strong}
\end{figure}

We evaluate strong scaling performance against cuSOLVERMp and MAGMA on 1, 2, 4, 8 GPU configurations for a $49,152 \times 49,152$ symmetric matrix in Figure~\ref{fig:strong}, and the numbers denote the speedups compared to the elapsed time on 1 GPU. As cuSOLVERMp cannot handle such matrix on one GPU, we use cuSOLVER's \texttt{Dsyevd} instead. As shown in Figure~\ref{fig:strong}, both benchmarks exhibit poor strong scalability on A100 GPUs: MAGMA and cuSOLVER fail to scale. On H100, MAGMA demonstrates negative strong scaling, exposing the limitations of hybrid CPU-GPUs implementations. cuSOLVERMp only exhibits marginal scaling, and it is probably because one-stage EVD has limited arithmetic intensity~\cite{williams2009roofline, DBBR}, which is hard to overlap data movements. In contrast, our pipelined EVD algorithm maintains robust scalability across both platforms. This consistent performance across hardware generations validates the effectiveness of our pipeline design.  

\begin{figure}
   \begin{subfigure}[b]{1.0\columnwidth}
    \includegraphics[width=\textwidth]{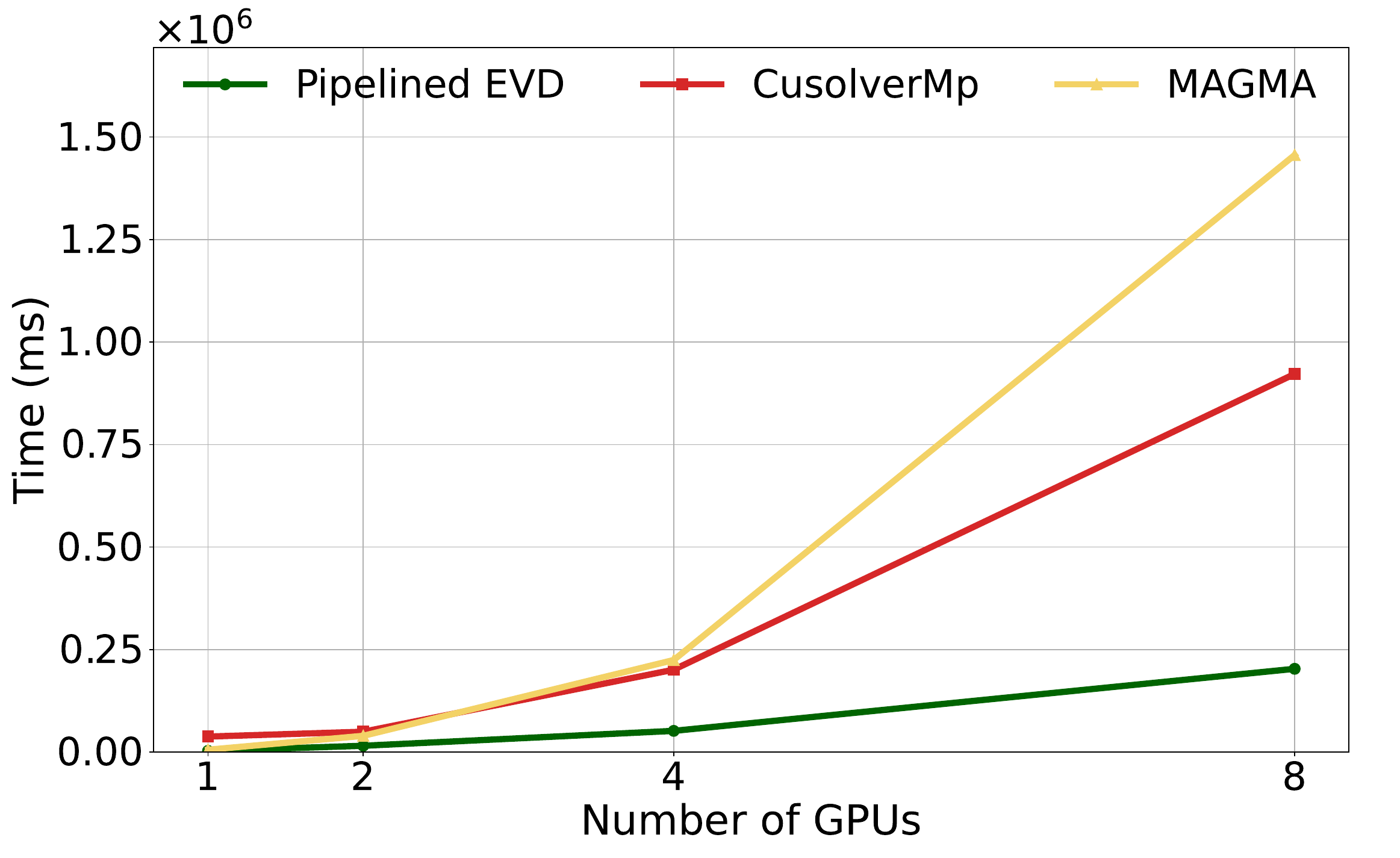}
    \caption{The weak scalability comparison on A100 GPUs.}
     \label{fig:a800_weak}
    \end{subfigure}
    
      \begin{subfigure}[b]{1.0\columnwidth}
    \includegraphics[width=\textwidth]{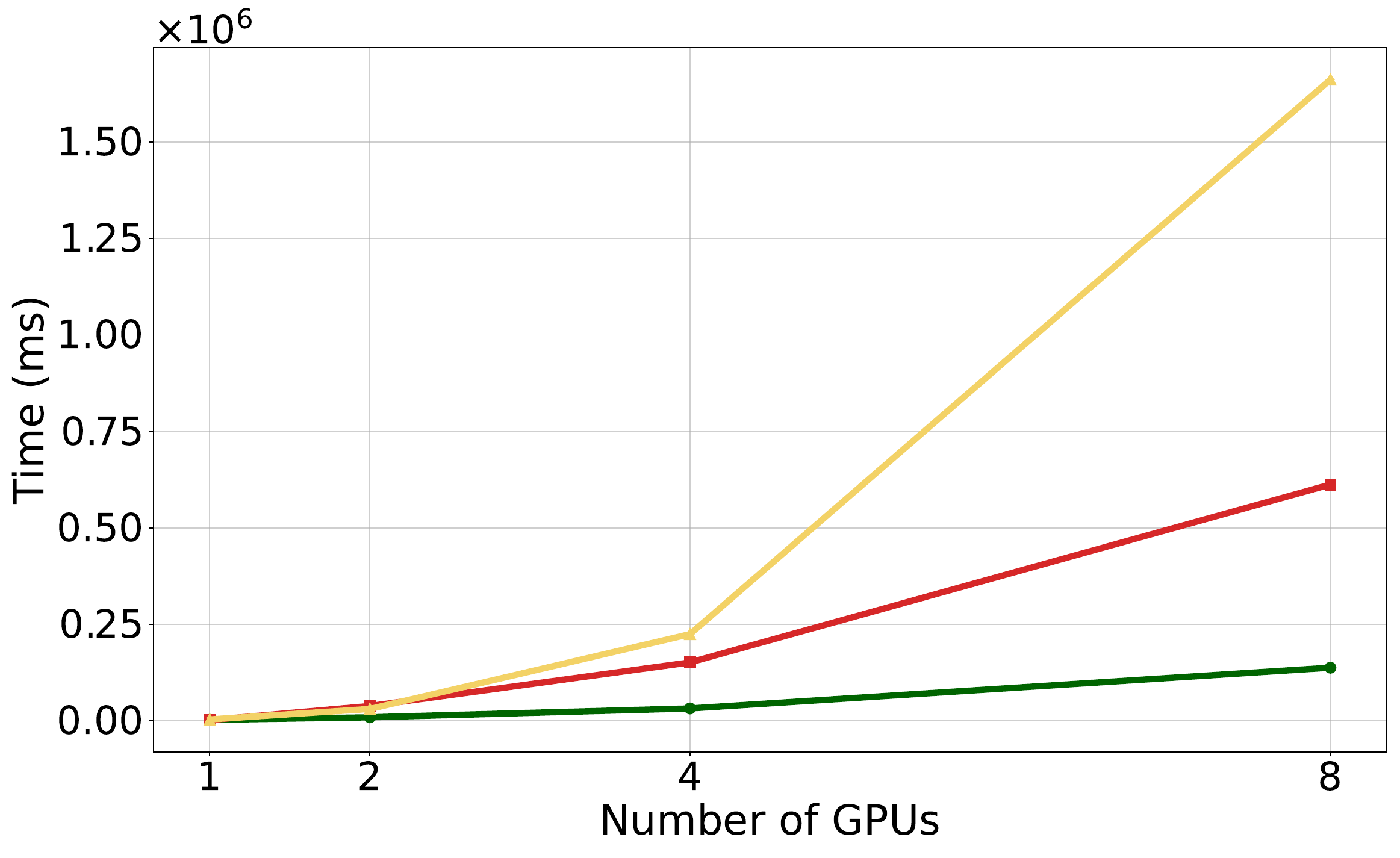}
    \caption{The weak scalability comparison on H100 GPUs.}
     \label{fig:h100_weak}
\end{subfigure}
\caption{The weak scalability comparison among cuSOLVERMp, MAGMA, and the our proposed pipelined EVD.}
\label{fig:weak}
\end{figure}

The weak scalability is illustrated in Figure~\ref{fig:weak}. The EVO solvers are executed across 1,2,4 and 8 GPUs with problem sizes scaled proportionally: 12288$\times$12288 (1 GPU), 24576$\times$24576 (2 GPUs), 49152$\times$49152 (4 GPUs), and 98304$\times$98304 (8 GPUs). Based on the results, our algorithm demonstrates much better weak Scalability than cuSOLVERMp and MAGMA, near-perfect weak scalability on both A100 and H100 platforms, confirming that the pipelined EVD delivers superior scalability for large-scale eigenvalue problems.

To sum up, the proposed pipelined EVD demonstrate better scalability than cuSOLVERMp and MAGMA. We think the there are two main reasons. First, pipelined EVD has less synchronization across different GPUs in the tridiagonalization process (SBR and BC). In other words, each GPU concentrates on its own task and minimizing inter-GPU synchronization. Second, the largest bottleneck in two-stage EVD is the BC-Back, and the BLAS2-based implementation provides less communication than BLAS3-based implementation, leading to better BC-back scalability (Figure~\ref{fig:line_bc_back}).

\subsection{Comparison with Block-Cyclic Implementation}

\begin{figure}
    \centering
    \includegraphics[width=1.0\columnwidth]{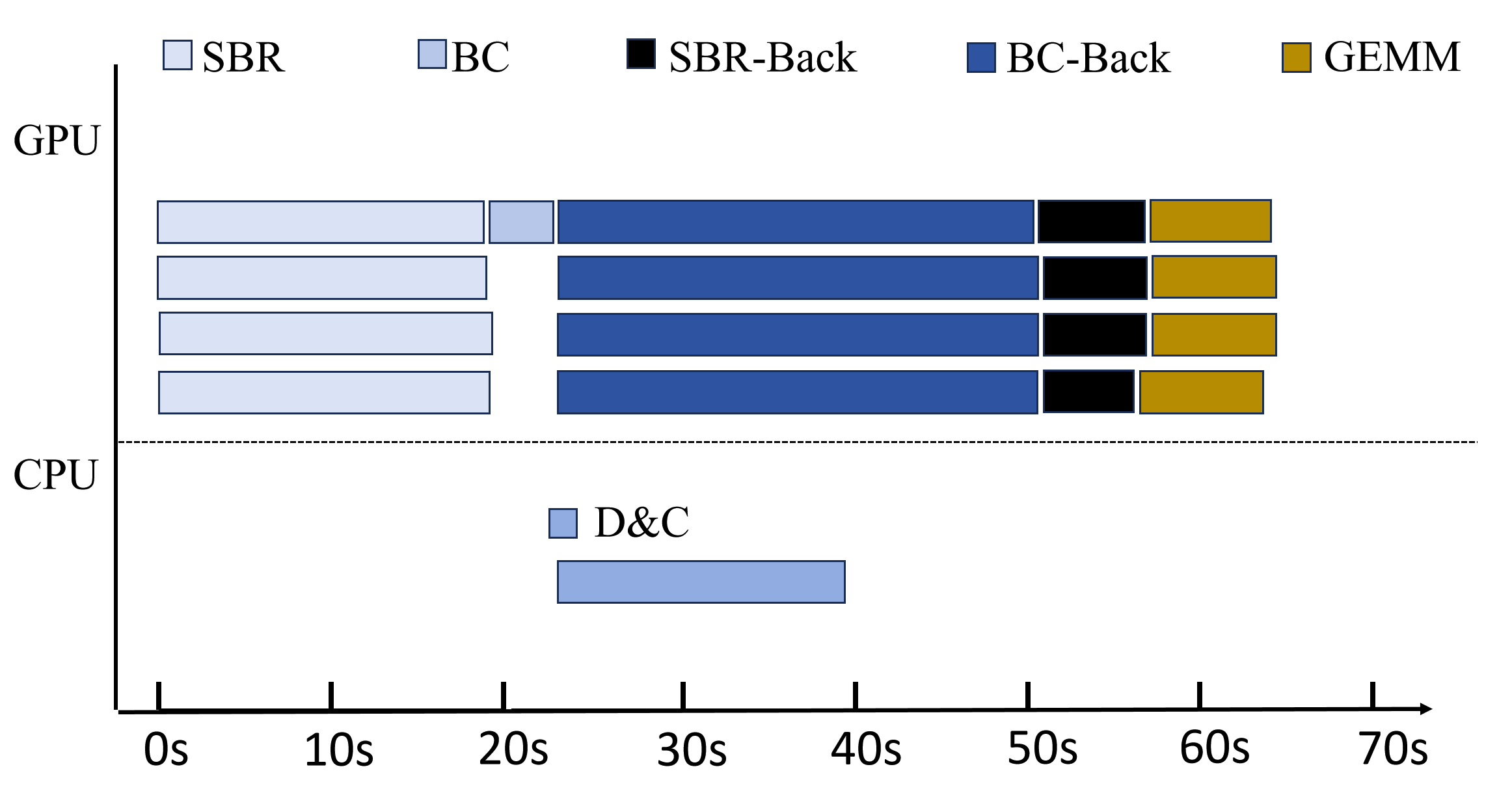}
    \caption{The timeline of block-cyclic EVD on a $49152\times 49152$ matrix using proposed optimizations on 4 A100 GPUs.\label{fig:line_bloccyclic}}
\end{figure}

To rigorously validate the efficiency of the pipelined EVD, we implement a block-cyclic EVD. Based on previous analysis, the block-cyclic strategy is difficult to leverage pipeline, so that the stages are executed in sequence. Compared to MAGMA's block-cyclic EVD, the new baseline involves the same optimizations as pipelined EVD uses, including communication avoiding SBR, BLAS2-based BC-Back and reordered stages.(We empirically demonstrate that reordering the stages yields better performance.) Figure~\ref{fig:line_bloccyclic} shows the timeline of block-cyclic baseline on 4 A100 GPUs with a $49152 \times 49152$ matrix. 

Compared to Figure~\ref{fig:new_pipeline}, the key distinction lies in the SBR and BC process. The SBR of block-cyclic baseline is more balanced, but it is longer than imbalanced pipelined SBR. This can be attributed to two reasons. 1) Block-cyclic SBR yields more communication: it always needs data movements across all GPUs, while pipelined SBR no longer requires communication from GPU$_i$ if GPU$_i$ has finished its own task. 2) Block-cyclic SBR cannot degrade to single GPU SBR on the last GPU, while pipelined SBR on the last GPU direct uses \texttt{syr2k} instead of GEMM. For the BC process, the pipelined EVD leverages back transformation to utilize all of the GPUs, while in block-cyclic BC, GPUs except GPU0 are idle and they're waiting for the finalization of BC on GPU0. These advantages enable the pipelined EVD to achieve 55.4s execution time versus 63.4s for the block-cyclic implementation, making the motivation of pipelined EVD more reasonable. 

\section{Related Work}

Computing symmetric/Hermitian eigenvalue problems has a long history. In 1980s, LAPACK~\cite{LAPACK} includes \texttt{syevd} routine using the blocking one-stage tridiagonalization method~\cite{blockReduction}, but it is limited on single CPU platform. Real world applications such as density functional theory problems~\cite{densityFunction} demand dense large EVD solvers with matrix size over 100k, which cannot be solved by one single CPU. Thus, in 1990s, scaLAPACK~\cite{blackford1997scalapack} and was proposed to solver larger EVD problems on distributed CPU architectures. Further, with the development of multicore architectures, PLASMA~\cite{PLASMA} and Eigen~\cite{EigenLibrary} emerges to utilize the computing capacity of multicore CPUs. Later in 2010s, the emerging GPUs demonstrate superior performance over CPUs on dense linear algebra problems, and MAGMA~\cite{MAGMA}  builds a bridge between CPU and GPU that it provides EVD solvers on hybrid CPU-GPU architectures. Nvidia also develops their own EVD solvers including cuSOLVER and cuSOLVERMp, which target on single GPU and multi-GPU architectures respectively.

The symmetric/Hermitian EVD solvers contain one-stage and two-stage EVD. one-stage EVD is more suitable for small matrices, because it time complexity is lower than two-stage EVD and most of the matrix data can fit into cache~\cite{blockReduction}.  two-stage EVD~\cite{2stage1,2stage3,2stage2,TCEVD} converts many BLAS2 operations in one-stage EVD, but it increases the time complexity from $O(4n^3)$ to $O(6n^3)$ because BC-Back contributes $2n^3$ extra computations~\cite{2stagesvd}. Fortunately, the increased BLAS3 operations compensates the growing complexity, and the recent research on two-stage tridiagonalization demonstrate over 10x speedup over cuSOLVER's one-stage tridiagonalization~\cite{DBBR}.

The applications of large symmetric/Hermitian EVD solvers mainly lie in Physics and Chemistry, and there are already some scientific computing software assembles EVD solvers, including VASP~\cite{vasp}, BerkeleyGW~\cite{berkeleygw} and Gromacs~\cite{van2005gromacs}.  It is noteworthy that these software typically adopts ELPA~\cite{marek2014elpa, ELPA2}, which support both one-stage and two-stage on multiple CPU-GPU architectures, as their built-in EVD solvers.

\section{Conclusion}
In this paper, we propose a pipelined EVD solver on multi-GPU architectures. The motivation of pipelining the EVD solver is that conventional two-stage EVD implementations do not have load balance and lack utilization of GPU resources, which can be solved by pipeline potentially.

However, there are some difficulties on converting the existing two-stage EVD solver to pipelined solver.
The first is that the conventional block-cyclic data distribution strategy does not allow as to pipeline SBR and BC, as BC demands the full GPU memory bandwidth and the communication between SBR and BC is intolerable. The second is that D\&C keeps updates the orthogonal matrix until the final conquer process, resulting in a pipeline stall. Therefore, to make the pipeline work, we change the distribution to blockwise distribution and then regard D\&C as an independent task, to maintain the pipeline without D\&C. Nevertheless, these modifications incur much more severe load imbalance, but we found it can be solved by adjusting the blocksize in SBR-Back and BC-Back, thereby balance the load in SBR and BC. In other words, we consider the four stages to be a whole task in terms of load balance. Experimentally, the proposed pipelined EVD solver utilizes the GPUs more efficiently. 

To further improve the performance, we propose several optimization techniques to reduce the communication in distributed SBR and BC. We also found the BLAS3-based BC-Back is inefficient because it increases the time complexity and the sizes and shapes of the GEMMs are too special to be executed efficiently. Thus, we use BLAS2-based BC-Back on multiple GPUs which is over 3.0x faster than conventional BC-Back implementation. Eventually, our implementation surpasses cuSOLVERMp and MAGMA baselines, delivering mean speedups of 5.74$\times$ and 6.59$\times$, respectively on A100 GPUs.

Our future work lies on enlarge the scale of EVD problems, as there are also some applications require decomposing $1M\times 1M$ symmetric/Hermitian matrices, which can only be solved on supercomputers with thousands of GPUs. Additionally, migrating the proposed techniques to non-symmetric matrices will be another challenge topic.

\bibliography{ref}

@article{QRAlgorithm,
  title={Understanding the QR algorithm},
  author={Watkins, David S},
  journal={SIAM review},
  volume={24},
  number={4},
  pages={427--440},
  year={1982},
  publisher={SIAM}
}

@book{MatrixComputation,
  title={Matrix computations},
  author={Golub, Gene H and Van Loan, Charles F},
  year={2013},
  publisher={JHU press}
}

@inproceedings{2stage1,
  title={Parallel reduction to condensed forms for symmetric eigenvalue problems using aggregated fine-grained and memory-aware kernels},
  author={Haidar, Azzam and Ltaief, Hatem and Dongarra, Jack},
  booktitle={Proceedings of 2011 International Conference for High Performance Computing, Networking, Storage and Analysis},
  pages={1--11},
  year={2011}
}

@incollection{2stage2,
  title={Solving the generalized symmetric eigenvalue problem using tile algorithms on multicore architectures},
  author={Ltaief, Hatem and Luszczek, Piotr and Haidar, Azzam and Dongarra, Jack},
  booktitle={Applications, Tools and Techniques on the Road to Exascale Computing},
  pages={397--404},
  year={2012},
  publisher={IOS Press}
}

@inproceedings{2stage3,
  title={Two-stage tridiagonal reduction for dense symmetric matrices using tile algorithms on multicore architectures},
  author={Luszczek, Piotr and Ltaief, Hatem and Dongarra, Jack},
  booktitle={2011 IEEE International Parallel \& Distributed Processing Symposium},
  pages={944--955},
  year={2011},
  organization={IEEE}
}

@book{LAPACK,
  title={LAPACK Users' guide},
  author={Anderson, Edward and Bai, Zhaojun and Bischof, Christian and Blackford, L Susan and Demmel, James and Dongarra, Jack and Du Croz, Jeremy and Greenbaum, Anne and Hammarling, Sven and McKenney, Alan and others},
  year={1999},
  publisher={SIAM}
}

@article{MAGMA,
  title={MAGMA Users’ Guide},
  author={Tomov, Stanimire and Nath, Rajib and Du, Peng and Dongarra, Jack},
  journal={ICL, UTK (November 2009)},
  year={2011}
}

@article{blockReduction,
  title={Block reduction of matrices to condensed forms for eigenvalue computations},
  author={Dongarra, Jack J and Sorensen, Danny C and Hammarling, Sven J},
  journal={Journal of Computational and Applied Mathematics},
  volume={27},
  number={1-2},
  pages={215--227},
  year={1989},
  publisher={Elsevier}
}

@article{2stagesvd,
  title={Accelerating the SVD two stage bidiagonal reduction and divide and conquer using GPUs},
  author={Gates, Mark and Tomov, Stanimire and Dongarra, Jack},
  journal={Parallel Computing},
  volume={74},
  pages={3--18},
  year={2018},
  publisher={Elsevier}
}

@article{williams2009roofline,
  title={Roofline: an insightful visual performance model for multicore architectures},
  author={Williams, Samuel and Waterman, Andrew and Patterson, David},
  journal={Communications of the ACM},
  volume={52},
  number={4},
  pages={65--76},
  year={2009},
  publisher={ACM New York, NY, USA}
}

@article{EigenLibrary,
  title={Eigen},
  author={Guennebaud, Ga{\"e}l and Jacob, Benoit and others},
  journal={URl: http://eigen. tuxfamily. org},
  volume={3},
  number={1},
  year={2010}
}

@article{DivideAndConquer,
  title={A parallel divide and conquer algorithm for the symmetric eigenvalue problem on distributed memory architectures},
  author={Tisseur, Fran{\c{c}}oise and Dongarra, Jack},
  journal={SIAM Journal on Scientific Computing},
  volume={20},
  number={6},
  pages={2223--2236},
  year={1999},
  publisher={SIAM}
}

@article{PLASMA,
  title={PLASMA: Parallel linear algebra software for multicore using OpenMP},
  author={Dongarra, Jack and Gates, Mark and Haidar, Azzam and Kurzak, Jakub and Luszczek, Piotr and Wu, Panruo and Yamazaki, Ichitaro and YarKhan, Asim and Abalenkovs, Maksims and Bagherpour, Negin and others},
  journal={ACM Transactions on Mathematical Software (TOMS)},
  volume={45},
  number={2},
  pages={1--35},
  year={2019},
  publisher={ACM New York, NY, USA}
}

@inproceedings{DBBR,
  title={Improving Tridiagonalization Performance on GPU Architectures},
  author={Wang, Hansheng and Duan, Zhekai and Zhao, Zitian and Wu, Siqi and Zheng, Saiqi and Li, Qiao and Jiang, Xu and Zhang, Shaoshuai},
  booktitle={Proceedings of the 30th ACM SIGPLAN Annual Symposium on Principles and Practice of Parallel Programming},
  pages={469--480},
  year={2025}
}

@book{blackford1997scalapack,
  title={ScaLAPACK users' guide},
  author={Blackford, L Susan and Choi, Jaeyoung and Cleary, Andy and D'Azevedo, Eduardo and Demmel, James and Dhillon, Inderjit and Dongarra, Jack and Hammarling, Sven and Henry, Greg and Petitet, Antoine and others},
  year={1997},
  publisher={SIAM}
}

@article{gates2020slate,
  title={SLATE users' guide},
  author={Gates, Mark and Charara, Ali and Kurzak, Jakub and YarKhan, Asim and Farhan, Mohammed Al and Sukkari, Dalal and Dongarra, Jack},
  year={2020}
}

@article{marek2014elpa,
  title={The ELPA library: scalable parallel eigenvalue solutions for electronic structure theory and computational science},
  author={Marek, Andreas and Blum, Volker and Johanni, Rainer and Havu, Ville and Lang, Bruno and Auckenthaler, Thomas and Heinecke, Alexander and Bungartz, Hans-Joachim and Lederer, Hermann},
  journal={Journal of Physics: Condensed Matter},
  volume={26},
  number={21},
  pages={213201},
  year={2014},
  publisher={IOP Publishing}
}

@article{tightBinding,
  title={Electromagnetic coupling in tight-binding models for strongly correlated light and matter},
  author={Li, Jiajun and Golez, Denis and Mazza, Giacomo and Millis, Andrew J and Georges, Antoine and Eckstein, Martin},
  journal={Physical Review B},
  volume={101},
  number={20},
  pages={205140},
  year={2020},
  publisher={APS}
}

@article{densityFunction,
  title={Density functional theory: a powerful tool for theoretical studies in coordination chemistry},
  author={Chermette, H},
  journal={Coordination chemistry reviews},
  volume={178},
  pages={699--721},
  year={1998},
  publisher={Elsevier}
}

@article{quantumChemistry1,
  title={Quantum chemical studies of light harvesting},
  author={Curutchet, Carles and Mennucci, Benedetta},
  journal={Chemical reviews},
  volume={117},
  number={2},
  pages={294--343},
  year={2017},
  publisher={ACS Publications}
}

@article{ELPA2,
  title={GPU-acceleration of the ELPA2 distributed eigensolver for dense symmetric and hermitian eigenproblems},
  author={Yu, Victor Wen-zhe and Moussa, Jonathan and Kus, Pavel and Marek, Andreas and Messmer, Peter and Yoon, Mina and Lederer, Hermann and Blum, Volker},
  journal={Computer Physics Communications},
  volume={262},
  pages={107808},
  year={2021},
  publisher={Elsevier}
}

@article{vasp,
  title={Ab-initio simulations of materials using VASP: Density-functional theory and beyond},
  author={Hafner, J{\"u}rgen},
  journal={Journal of computational chemistry},
  volume={29},
  number={13},
  pages={2044--2078},
  year={2008},
  publisher={Wiley Online Library}
}

@article{berkeleygw,
  title={BerkeleyGW: A massively parallel computer package for the calculation of the quasiparticle and optical properties of materials and nanostructures},
  author={Deslippe, Jack and Samsonidze, Georgy and Strubbe, David A and Jain, Manish and Cohen, Marvin L and Louie, Steven G},
  journal={Computer Physics Communications},
  volume={183},
  number={6},
  pages={1269--1289},
  year={2012},
  publisher={Elsevier}
}

@article{van2005gromacs,
  title={GROMACS: fast, flexible, and free},
  author={Van Der Spoel, David and Lindahl, Erik and Hess, Berk and Groenhof, Gerrit and Mark, Alan E and Berendsen, Herman JC},
  journal={Journal of computational chemistry},
  volume={26},
  number={16},
  pages={1701--1718},
  year={2005},
  publisher={Wiley Online Library}
}

@inproceedings{TCEVD,
  title={Fast symmetric eigenvalue decomposition via wy representation on tensor core},
  author={Zhang, Shaoshuai and Shah, Ruchi and Ootomo, Hiroyuki and Yokota, Rio and Wu, Panruo},
  booktitle={Proceedings of the 28th ACM SIGPLAN Annual Symposium on Principles and Practice of Parallel Programming},
  pages={301--312},
  year={2023}
}
\end{document}